\documentstyle[eqsecnum,preprint,aps,prd,mathrsfs]{revtex}
\def\bw{\bar\omega}
\def\ba{\bar\alpha}
\def\bb{\bar\beta}
\def\half{\frac{1}{2}}
\def\ts{\tilde s}
\def\sr{s_{R}}
\def\dl{\delta\lambda}
\def\slash#1{\, /\kern-0.6em{#1}}

\begin{document}
\draft
\tightenlines

\preprint{\vbox{\hbox{hep-th/9911107}\smallskip\hbox{SBNC/99-11-01}}}


\title{Renormalizability of the Dynamical Two-Form}     
\author{Amitabha Lahiri}
\address{S. N. Bose National Centre for Basic Sciences, \\
Block JD, Sector III, Salt Lake, Calcutta 700 091, INDIA}
\address{amitabha@boson.bose.res.in}
\date{Physical Review D (to be published)}
\maketitle

\begin{abstract}

A proof of renormalizability of the theory of the dynamical
non-Abelian two-form is given using the Zinn-Justin equation. Two
previously unknown symmetries of the quantum action, different from
the BRST symmetry, are needed for the proof. One of these is a gauge
fermion dependent nilpotent symmetry, while the other mixes different
fields with the same transformation properties. The BRST symmetry
itself is extended to include a shift transformation by use of an
anticommuting constant. These three symmetries restrict the form of
the quantum action up to arbitrary order in perturbation theory. The
results show that it is possible to have a renormalizable theory of
massive vector bosons in four dimensions without a residual Higgs
boson.

\end{abstract}

\medskip


\section{Introduction}
Each and every aspect of the Standard Model has been tested in recent
years, with remarkable agreement with theory, except in one
sector. The Standard Model predicts the existence of the Higgs boson,
responsible for making gauge bosons and fermions massive, as well as
breaking the SU(2)$\times$U(1) symmetry of the theory down to the U(1)
of electromagnetism. But no elementary scalar has yet been observed in
any particle interaction, nor has any experiment so far detected the
Higgs boson, either elementary or composite. On the other hand,
various theoretical constraints put the upper bound of the Higgs boson
mass only a little out of reach of present day experiments. It is
therefore useful to consider the scenario in which the Higgs boson
remains unobserved as the theoretical bounds are reached.

Apart from the Higgs boson and a possible neutrino mass, the Standard
Model agrees quite closely with experiment, so it is a good idea to
leave most of the theory untouched. The role of the Higgs boson may be
distributed among possibly different mechanisms for generating vector
and fermion masses, and symmetry breaking. The Higgs mechanism does
all this in a renormalizable and unitary way \cite{thooft}, and any
alternative must not affect these good quantum properties of the
theory.  A possible alternative for generating vector boson masses is
to use a dynamical two-form. When an antisymmetric tensor potential
$B$ is coupled to the field strength $F$ of a U(1) gauge field via a
`topological' $B\wedge F$ coupling and a kinetic term for $B$ is
included, the gauge field develops an effective
mass\cite{aurtak,trg,abl,minwar}. The mass is equal to the
dimensionful coupling constant $m$ of the interaction term, and there
is no residual scalar (Higgs) degree of freedom. If a non-Abelian
version of this theory can be consistently quantized, it may be
applied to particle interactions.

No-go theorems \cite{hurth,nogo}\, based on the consistency of
quantum symmetries rule out most, but not all, alternative Higgs
free mechanisms of mass generation for non-Abelian vector
bosons. One useful exception is the topological mass generation
mechanism \cite{gvm}\, which has seen renewed interest in recent
years \cite{hwalee,nabrst,neto}. This mechanism uses an auxiliary
vector field to close the symmetry algebra and thus avoid the no-go
theorems. The price one has to pay is to have non-propagating
bosonic and ghost fields in the theory, which disappear in the
Abelian limit.  The no-go theorem of Ref.~\cite{nogo}\, says only
that the non-Abelian model cannot be constructed from the Abelian
model, which is known to be quantizable \cite{abl}. It does not
rule out the quantizability of the non-Abelian model itself.
However that is not in itself a proof that the non-Abelian model is
quantizable, and a proof has not been constructed as yet. The first
step in such a proof is the construction of a BRST-invariant
tree-level action, which was done from a geometric point of view in
\cite{hwalee} and ab initio in \cite{nabrst}.

In this paper I construct the quantum action for this model up to
arbitrary order in perturbation theory starting from the
BRST-invariant tree-level action. I follow an algebraic procedure
along the lines of what is done for Yang-Mills theories
\cite{ZJ,weinqft}. The construction itself is rather involved as
there are different fields with the same transformation
properties. This suggests that the usual BRST symmetry is not
sufficient to restrict the operators in the quantum
action. Fortunately there are other useful symmetries of the
tree-level action and they, together with the BRST symmetry, are
sufficient for the purpose. The starting point of the paper is the
classical action given in Sec.~\ref{lag}. In Sec.~\ref{brs}\, I
list the BRST transformation rules of the theory and construct
another BRST-like nilpotent symmetry.  In Sec.~\ref{sym}\, I
construct the quantum symmetries corresponding to these and other
symmetries, and in Sec.~\ref{ren}\, I find all the dimension four
operators allowed by all the symmetries. Finally, Sec.~\ref{disc}\,
carries a small discussion of possible extensions and applications
of the results. The main body of the paper sets up the structure of
the proof, most of the detailed calculations are collected in
appendices at the end.

\section{Tree level action}\label{lag}
In this section I shall fix my conventions. I shall work with an
SU(N) gauge group $G$, with generators $t_a$ satisfying 
\begin{equation}
\left[ t_a, t_b \right] = if^{abc}\,t_c,
\end{equation}
with the structure constants $f^{abc}$ totally antisymmetric in its
indices. The gauge index, as well as Lorentz indices, will be made
explicit in general for easier tracking of numerical
coefficients. The background metric is taken to have signature
$(-+++)$. 

The classical action for the dynamical non-Abelian two-form
\cite{gvm} is
\begin{equation}
S_0 = \int d^4x \bigg(- {1\over 4}F_{\mu\nu}^aF^{a\mu\nu} - {1\over
12}H_{\mu\nu\lambda}^aH^{a\mu\nu\lambda} + {m\over 4}
\epsilon^{\mu\nu\rho\lambda}B_{\mu\nu}^aF_{\rho\lambda}^a\bigg) \,.
\label{brs.S0}
\end{equation}
Here $F_{\mu\nu}$ is the curvature of a gauge connection $A_\mu$
with gauge coupling $g$,
\begin{equation}
F_{\mu\nu}^a = ({i\over g}[D_\mu, D_\nu])^a = \partial_\mu A_\nu^a
- \partial_\nu A_\mu^a + gf^{abc}A_\mu^b A_\nu^c\,.
\label{brs.F}
\end{equation}
The compensated field strength $H_{\mu\nu\lambda}$ is defined with
the help of an auxiliary field $C_\mu$ by the relation
\begin{eqnarray}
H_{\mu\nu\lambda}^a &=& (D_{[\mu}B_{\nu\lambda]})^a +
ig\left[
F_{[\mu\nu}, C_{\lambda]}\right]^a \nonumber\\
&=& \partial_{[\mu}B_{\nu\lambda]}^a +
gf^{abc}A^b_{[\mu}B^c_{\nu\lambda]}
- gf^{abc}F_{[\mu\nu}^bC_{\lambda]}^c.
\label{brs.hdef}
\end{eqnarray}
All the three fields $A_\mu$, $B_{\mu\nu}$ and $C_\mu$ belong to
the adjoint representation of the gauge group $G$.  The action
(\ref{brs.S0}) therefore remains invariant under gauge
transformations given by
\begin{eqnarray}
A_\mu \to U A_\mu U^{-1} - {i\over g}\partial_\mu UU^{-1}, \qquad
B_{\mu\nu} \to U B_{\mu\nu} U^{-1},  
\qquad C_\mu \to U C_\mu U^{-1}, \qquad U\in G\,.
\label{brs.gxfn}
\end{eqnarray}
In addition, the action $S_0$ is invariant under vector
gauge transformations given by
\begin{eqnarray}
A_\mu \to A_\mu, \qquad B_{\mu\nu} \to B_{\mu\nu} +
D_{[\mu}\Lambda_{\nu]}, \qquad C_\mu \to C_\mu + \Lambda_\mu\,,
\label{brs.vxfn}
\end{eqnarray}
where $\Lambda_\mu$ is some arbitrary vector field in the adjoint
representation of the gauge group which vanishes at infinity.

For the purpose of power counting, I need the propagators of this
theory. Let me choose the usual Lorenz gauge $\partial_\mu A^{a\mu}
= 0$, $\partial_\nu B^{a\mu\nu} = 0$, with gauge parameters $\xi$
and $\eta$ respectively. Then the tree-level propagator for
$A^a_\mu$ is
\begin{eqnarray}
D^{ab}_{\mu\nu} = - {\delta^{ab}\over k^2}\left(g_{\mu\nu} -
\left(1 - \xi \right){k_\mu k_\nu \over k^2} \right),
\end{eqnarray}
and that for $B^a_{\mu\nu}$ is 
\begin{eqnarray}
D^{ab}_{\mu\nu\,,\,\rho\lambda} = - {\delta^{ab}\over
k^2}\left(g_{\mu[\rho}g_{\lambda]\nu} 
- \left(1 - \eta \right){g_{\mu[\rho}k_{\lambda]} k_\nu 
\over k^2}\right)\,. 
\end{eqnarray}

There is no quadratic term in the action involving the auxiliary
field $C^a_\mu$ in this gauge, so the tree level propagator for it
will vanish. As a result, there are no diagrams with internal
$C^a_\mu$ lines. This may look very peculiar, but it is also
possible to choose some other gauge for which $C^a_\mu$ has a
non-vanishing (gauge-dependent) propagator, but that does not
change the arguments. This is discussed in the last section. For
the moment let me proceed without a quadratic term for the
auxiliary field $C^a_\mu$.


There is a quadratic coupling term between the vector and the
antisymmetric tensor fields coming from the last term in the
action, the vertex given by
\begin{eqnarray}
V^{ab}_{\nu,\,\rho\lambda} = im\delta^{ab}
\epsilon_{\mu\nu\rho\lambda} k^\mu\,.
\end{eqnarray}
The effective tree-level propagator for the vector field is then
calculated by summing all insertions of the tree-level $B_{\mu\nu}$
propagator into the na\"\i ve tree-level $A_\mu$ propagator
\cite{abl}. The result is 
\begin{eqnarray}
\widetilde D^{ab}_{\mu\nu} = - {\delta^{ab}\over k^2 -
m^2}\left(g_{\mu\nu} - {k_\mu k_\nu \over k^2}\right) -
\delta^{ab}\xi\,{k_\mu k_\nu \over k^4 } \,.
\end{eqnarray}
This shows that there is a pole in the two-point function of the
vector field $A^a_\mu$ even at tree-level. On the other hand, the
`massive' propagator $\widetilde D^{ab}_{\mu\nu}$ falls off as
$1/k^2$ at high values of $k^2$, like in the case of the Higgs
mechanism. The ultraviolet behavior of the propagators show that
the theory is power-counting renormalizable in this gauge. The best
way to proceed further is via the BRST method of quantization.

\section{BRST Invariance}\label{brs}
Quantization of this theory requires gauge-fixing and therefore the
introduction of ghosts. The gauge fixed action, together with the
ghost terms, is BRST invariant. The vector gauge symmetry requires
ghosts of ghosts, and off-shell nilpotence of the BRST charge
requires auxiliary fields. Let me write the gauge-fixing functions
as $f^a$, $f^{a\mu}$ and $f'^a$ for gauge transformations, vector
gauge transformations and gauge transformations of ghosts,
respectively. In Sec.~\ref{ren}, I shall choose the gauge functions
to be of the usual Lorenz gauge type,
\begin{equation}
f^a = \partial_\mu A^{a\mu}, \qquad f^{a\mu} = \partial_\nu
B^{a\mu\nu}, \qquad f'^a = \partial_\mu\omega^{a\mu}\,,
\label{brs.gauges}
\end{equation}
but most of the results in this paper will hold for arbitrary
linear gauge functions. Some discussion about arbitrary gauges is
presented in Sec.~\ref{disc}.

The tree level quantum action can be written as
\begin{eqnarray}
S = S_0 + \int d^4x &&\left[ h^a f^a  + \bw^a \Delta^a +
\half\xi h^a h^a + \right. 
h^a_\mu \left( f^{a\mu} + \partial^\mu n^a
\right) 
\nonumber \\
&& + \bw^a_\mu \Delta^{a\mu} + \half\eta h^a_\mu h^{a\mu} 
-  \left.\partial_\mu \bw^{a\mu}\alpha^a + \ba^a f'^a 
+ \bb^a \Delta'^a + \zeta\ba^a\alpha^a \right]\,.
\label{brs.treeS}
\end{eqnarray}
Here $\Delta^a$, $\Delta^{a\mu}$ and $\Delta'^a$ are the BRST
variations, as defined below, of $f^a$, $f^{a\mu}$ and $f'^a$,
respectively. The appearance of $\partial_\mu n^a$ in the gauge-fixing
condition is usual for two-form gauge-fields. The gauge-fixing
condition $f^{a\mu} = 0$ holds upon using the equation of motion of
$n^a$ \cite{hentei}.  This action is no longer invariant under gauge
or vector gauge transformations. But it is invariant under the BRST
transformations  \cite{hwalee,nabrst}
\begin{eqnarray}
&s& A^a_\mu = \partial_\mu \omega^a + g f^{abc}A^b_\mu \omega^c
\,,\nonumber \\
&s& \omega^a = - {1\over 2}
gf^{abc}\omega^b\omega^c \,, \qquad 
s \bw^a = - h^a \,, \qquad 
s h^a = 0 \,,\nonumber \\
&s& B^a_{\mu\nu} = gf^{abc}B^b_{\mu\nu}\omega^c +
(D_{[\mu}\omega_{\nu]})^a
+ gf^{abc}F^b_{\mu\nu}\theta^c\,,\nonumber\\
&s& C^a_\mu = gf^{abc}C^b_\mu\omega^c + \omega^a_\mu +
(D_\mu\theta)^a\,,\nonumber\\
&s& \omega^a_\mu = -gf^{abc}\omega^b_\mu\omega^c +
(D_\mu\beta)^a \,,\nonumber\\
&s& \bw^a_\mu = - h^a_\mu \,,\qquad 
s h^a_\mu = 0 \,, \qquad 
s n^a = \alpha^a \,, \qquad 
s \alpha^a = 0 \,,\nonumber\\
&s& \beta^a = gf^{abc}\beta^b\omega^c  \,,\nonumber\\
&s& \bb^a = \ba^a \,, \qquad 
s \ba^a = 0 \,,\nonumber \\
&s& \theta^a = - gf^{abc}\theta^b\omega^c - \beta^a
\,.
\label{brs.brst}
\end{eqnarray}
These transformations are nilpotent, $s^2 = 0$ on all fields, if
$s$ has a left action, i.e., the change in any field $\chi^A$ is
given by $\delta\chi^A = \delta\lambda s\chi^A$, where
$\delta\lambda$ is an anticommuting infinitesimal parameter. The
tree-level quantum action of Eqn.(\ref{brs.treeS}) is invariant
under $s$, with $\Delta^a = sf^a$, $\Delta^{a\mu} = sf^{a\mu}$ and
$\Delta'^a = sf'^a$. It is also possible to write this action as the
sum of the classical action $S_0$ plus a total super-divergence,
\begin{eqnarray}
S &=& S_0 + s \Psi\,, \qquad{\rm with} \nonumber \\
\Psi = -\left( \bw^a f^a + \half\xi \bw^a h^a \right) 
&-& \left( \bw^a_\mu \left( f^{a\mu} + \partial^\mu n^a \right) 
+ \half \eta \bw^a_\mu h^{a\mu} \right) 
+ \left( \bb^a f'^a + \zeta \bb^a \alpha^a \right) \,.
\label{brs.Psi}
\end{eqnarray}

In addition to the BRST transformations, there is another BRST-type
nilpotent transformation which leaves the action invariant. Such a
symmetry exists for all gauge theories, not just the two-form
theories, as can be seen from the following argument. The terms in
the extended ghost sector of the tree-level quantum action
of a gauge theory are typically of the form
\begin{eqnarray}
S^c_{ext} =   h^A f^A + \half \lambda h^A h^A +
\bw^A \Delta^A \,,
\label{brs.Scext}
\end{eqnarray}
where $(\bw^A, h^A)$ are the trivial pairs. Here the index $A$ stands
for the collection of various indices as well as the space-time point
where the fields are evaluated, $f^A = 0$ is the corresponding
gauge-fixing condition with gauge parameter $\lambda$, and $\Delta^A =
sf^A$. The sum over $A$ includes the integration over space-time. This
form of the extended ghost sector is valid for commuting $h^A$, $f^A$
and anticommuting $\bw^A$. For example, all but the last three terms
of the tree-level quantum action (\ref{brs.treeS}) can be written in
this form, where the index $A$ includes the gauge index $a$ or the
pair $(a,\mu)$ depending on the gauge field ($A_\mu$ or
$B_{\mu\nu}$). This part of the action remains invariant under BRST
transformations
\begin{eqnarray}
s \bw^A = -h^A\,, \qquad s h^A = 0 \,.
\label{}
\end{eqnarray}
On the other hand, I can rearrange $S^c_{ext}$ as
\begin{eqnarray}
S^c_{ext} &=& \half \lambda \left( h^A + {1\over \lambda}
f^A\right) \left( h^A + {1\over \lambda} f^A\right) - {1\over
2\lambda} f^A f^A + \bw^A \Delta^A \nonumber \\ 
&=& \half \lambda \left( \left(h^A + {2\over \lambda} f^A \right) -
{1\over \lambda} f^A\right) \left( \left( h^A + {2\over \lambda}
f^A\right) - {1\over \lambda} f^A\right) - {1\over 2\lambda} f^A
f^A + \bw^A \Delta^A \nonumber \\
&=& \half \lambda \left( h'^A + {1\over \lambda} f^A\right)
\left( h'^A + {1\over \lambda} f^A\right) - {1\over 2\lambda} f^A
f^A + \bw^A \Delta^A \nonumber \\
&=& h'^A f^A + \half \lambda h'^A h'^A + \bw^A \Delta^A \,,
\label{brs.newh}
\end{eqnarray}
where I have defined $h'^A = - h^A - \displaystyle{2\over \lambda}
f^A$. So far, I have not actually done anything.
The only thing that comes out of this exercise is the fact that
$S^c_{ext}$ is invariant under a new set of BRST transformations:
\begin{eqnarray}
\ts \bw^A = - h'^A &\Rightarrow& \ts\bw^A = h^A + {2\over\lambda}
f^A \,,\nonumber \\ 
\ts h'^A  = 0 &\Rightarrow& \ts h^A = - {2\over\lambda} \ts f^A \,,
\nonumber \\ 
\ts f^A &=& \Delta^A \equiv s f^A\,.
\label{brs.cstilde}
\end{eqnarray}
Therefore, if the action of $\ts$ on $\bw^A$ and $h^A$ is as above,
and $\ts = s$ on all other fields, the last equation is identically
satisfied, and it also follows that $\ts$ is nilpotent on all
fields, $\ts^2 = 0$ if $\Delta^A$ does not contain any auxiliary
field, which is usually the case.

When the extended sector corresponds to an anticommuting gauge
field, as in the case of gauge-fixing of ghost fields, the
construction is slightly more complicated, since the auxiliary
fields have odd ghost number. Typically, for anticommuting
auxiliary fields $\ba^A$, $\alpha^A$, the extended ghost sector can
be written as 
\begin{eqnarray}
S^a_{ext} = \ba^A f'^A + \bar f'^A \alpha^A + \zeta \ba^A \alpha^A
+ \bb^A \Delta'^A \,.
\label{brs.Saext}
\end{eqnarray}
In this, $f'^A$ is the anticommuting gauge-fixing function,
$\Delta'^A = s f'^A$, and $\bb^A$ is the corresponding commuting
antighost. The term $\bar f'^A \alpha^A$ is just a rearrangement of
the appropriate terms in $\bw^A \Delta^A$ which appear for the
usual gauge symmetries. A term such as $\bar f'^A \alpha^A$ must
appear, since $\alpha^A$ is the ghost for some field and therefore
appears in some $\Delta^A$. For example, in the tree-level quantum
action of Eqn.(\ref{brs.treeS}) $\bar f'^A \alpha^A$ corresponds to
$ \bw^{a\mu}\partial_\mu
\alpha^a$, which in turn is required to cancel the BRST variation
of $h^a_\mu \partial^\mu n^a$. Just as in the case with commuting
auxiliary fields, the terms in $S^a_{ext}$ can be rearranged,
\begin{eqnarray}
S^a_{ext} &=& \zeta \left( \ba^A + {1\over \zeta} \bar f'^A \right)
\left( \alpha^A + {1\over \zeta}  f'^A \right) -  {1\over \zeta}
\bar f'^A f^A + \bb^A \Delta'^A \, \nonumber \\
&=& \zeta \left( \left( \ba^A + {2\over \zeta} \bar f'^A \right) -
{1\over \zeta} \bar f'^A \right)  \left( \left( \alpha^A + {2\over
\zeta} f'^A \right) - {1\over \zeta}  f'^A \right) -  {1\over
\zeta} \bar f'^A f^A + \bb^A \Delta'^A \, \nonumber \\
&=& \zeta \left( \ba'^A + {1\over \zeta} \bar f'^A \right)
\left( \alpha'^A + {1\over \zeta}  f'^A \right) -  {1\over \zeta}
\bar f'^A f^A + \bb^A \Delta'^A \, \nonumber \\
&=& \zeta \ba'^A \alpha'^A + \ba'^A f'^A + \bar f'^A \alpha'^A +
\bb^A \Delta'^A \,.
\label{brs.newalpha}
\end{eqnarray}
where I have now defined $\ba'^A = - \left( \displaystyle \ba^A +
{2\over \zeta}\bar f'^A \right)$ and $\alpha'^A = - \left(
\alpha^A+ \displaystyle {2\over \zeta} f'^A \right)$.  As
before, in these coordinates $S^a_{ext}$ is invariant under its own
set of BRST transformations,
\begin{eqnarray}
\ts \bb^A = \ba'^A = - \left( \ba^A + {2\over \zeta} \bar f'^A
\right) \,, \nonumber \\
\ts \ba'^A = 0 \Rightarrow \ts \ba^A = - {2\over \zeta} \ts \bar f'^A
\,, \nonumber \\
\ts \alpha'^A = 0 \Rightarrow \ts \alpha^A = - {2\over \zeta} \ts f'^A
\,, \nonumber \\
\ts f'^A = \Delta'^A \equiv s f'^A \,.
\label{brs.astilde}
\end{eqnarray}
Two more things are required for the nilpotence of $\ts$ ---
$\alpha^A$ was the result of BRST variation of some field
($\alpha^a = sn^a$ in Eqn.(\ref{brs.brst})) --- now $\alpha'^A$ has
to be the variation under $\ts$ of the same field, and $\ts \bar
f'^A$ must be calculated according to the rules of
Eqn.(\ref{brs.cstilde}) for $\ts$ acting on the anticommuting
ghosts in $\bar f'^A$. In addition, the action of $\ts$ must be the
same as that of $s$ for the fields contained in $f'^A$. Then $\ts^2
= 0$ on all fields.

I can now gather the results of Eqn.(\ref{brs.cstilde}) and
Eqn.(\ref{brs.astilde}) and apply them to the tree-level quantum
action of Eqn.(\ref{brs.treeS}) to construct this symmetry,
\begin{eqnarray}
\ts \bw^a &=&  h^a + {2\over \xi} \, f^a \,, \nonumber \\ \ts h^a
&=& -{2\over \xi} \Delta^a \,, \nonumber \\
\ts \bw^a_\mu &=& h^a_\mu + {2\over \eta} \, \partial_\mu n^a + 
{2\over \eta} \, f^a_\mu \,, \nonumber \\ \ts h^a_\mu &=& {2\over
\eta} \left( \partial_\mu \alpha^a + {2\over \zeta} \, \partial_\mu
f'^a - \Delta^a_\mu \right) \,, \nonumber \\
\ts n^a &=& - \left( \alpha^a + {2\over \zeta} \, f'^a \right) \,,
\nonumber \\ 
\ts \bb^a &=& - \left( \ba^a - {2\over \zeta} \,
\partial_\mu \bw^{a\mu} \right)\,, \nonumber \\ 
\ts \alpha^a &=& - {2\over \zeta} \Delta'^a \,, \nonumber \\
\ts \ba^a &=& {2\over \zeta}\, \ts \left( \partial_\mu
\bw^{a\mu}\right) = {2\over \zeta}\, \partial_\mu \left( h^{a\mu}
+ {2\over \eta } \, \partial^\mu n^a + {2\over \eta} f^{a\mu}
\right) \,, 
\label{brs.stilde} \\
\ts &=& s \mbox{ on all other fields.} \nonumber
\end{eqnarray}

Since the gauge-fixing functions do not contain antighosts or
auxiliary fields, and since BRST variations of the remaining fields
also do not contain antighosts or auxiliary fields, a
straightforward calculation shows that $\ts$ is nilpotent on all
fields,
\begin{eqnarray}
\ts^2 = 0\,. 
\end{eqnarray}
In addition, since the classical action $S_0$ is invariant under
BRST transformations, and since $\ts = s$ on the fundamental
fields, 
\begin{eqnarray}
\ts S_0 = 0\,.
\end{eqnarray}
The remainder of the tree-level quantum action of
Eqn.~(\ref{brs.treeS}) can be written as a sum of $S^c_{ext}$ and
$S^a_{ext}$ as defined above, and either by the method described
above or by an explicit calculation it can be shown quite easily
that this part is also invariant under $\ts$. So in fact
\begin{eqnarray}
\ts S = 0 \,.
\end{eqnarray}

It should be made clear that $\ts$ is not special to the dynamical
two-form, nor even to reducible gauge systems. Usual gauge theories
exhibit invariance under a symmetry analogous to $\ts$. But in
those cases, this gauge-fermion dependent invariance is not needed
for restricting the form of the quantum action --- invariance under
the familiar BRST transformation $s$ is sufficient for that purpose
\cite{ZJ,weinqft}. However, $\ts$ becomes extremely useful
 when the theory contains many different fields in the same
representation, as in the case of the dynamical two-form. I shall
make extensive use of $\ts$ to construct the quantum effective
action for the dynamical two-form. In order to do that, I need to
look at the quantum symmetries corresponding to $s, \ts$ and some
other classical symmetries of the theory. This is done in the next
section.

\section{Symmetries of the Effective Action}\label{sym}
On the way to a proof of perturbative renormalizability of the
dynamical non-Abelian two-form, the first thing to note is that
there is no kinetic term for $C^a_\mu$ in the tree-level action.
Consequently, $C^a_\mu$ is taken to be dimensionless.  The
auxiliary ghost field $\theta$ is taken to be dimensionless for the
same reason, and then the theory is power-counting renormalizable.
The presence of fields with vanishing mass dimension does not
automatically rule out renormalizability of a theory\cite{piso},
but it is possible that the theory will be non-renormalizable
because counterterms may contain arbitrary powers of these
fields. Therefore, one needs to ensure that the symmetries of the
theory restrict the number of counterterms to a finite value. 
Perturbative renormalizability requires that the quantum effective
action, invariant under the quantum symmetries, contain only those
operators which appear in the tree-level action of
Eqn.\,(\ref{brs.treeS}) up to arbitrary numerical coefficients. The
quantum action can be constructed by use of the Zinn-Justin
equation in the following manner.

The partition function $Z[J,K]$ in the presence of external
$c$-number sources $J^A(x), K^A(x)$ is 
\begin{eqnarray}
Z[J,K] = \int \left[{\mathscr D}\chi^B \right] \exp \left( iS + i\int
d^4x \chi^A J^A + i\int d^4x F^A K^A \right) \,,
\label{sym.newZ}
\end{eqnarray}
where $F^A(x) = s\chi^A(x)$, and I have kept the space-time
integration explicit for this section. I shall also refer to $K^A$
as the `antisource' corresponding to the field $\chi^A$. This
partition function leads to the effective action
\begin{eqnarray}
\Gamma[\chi, K] = - \int d^4x \chi^A J^A_{\chi,K} - i\ln
Z[J_{\chi,K},K] \,,
\label{sym.neweffac}
\end{eqnarray}
where $J^A_{\chi,K}$ is the value of the current for which
$\langle \chi^B(x)\rangle_{J,K} = \chi^B(x)$, the expectation value
being calculated in the presence of $K^A$. 

The effective action satisfies the Zinn-Justin equation
\cite{ZJ,weinqft}, 
\begin{eqnarray}
(\Gamma, \Gamma) = 0 \,,
\label{sym.ZJ}
\end{eqnarray}
where the antibracket $(F,G)$ is defined for any two functionals
$F$ and $G$ as
\begin{eqnarray}
(F,G) = \int d^4x\; {\delta_R F[\chi,K]\over \delta
\chi^A(x)}{\delta_L G[\chi,K]\over \delta K^A(x) } -
\int d^4x\; {\delta_R F[\chi,K]\over \delta
K^A(x)}{\delta_L G[\chi,K]\over \delta \chi^A(x) } \,.
\end{eqnarray}

In order to get a proof of perturbative renormalizability of a
theory, the total action functional $S[\chi,K] = S[\chi] + \int
d^4x F^A K^A$ is written as a sum of the renormalized action
$S_R[\chi, K]$ plus a term $S_\infty[\chi,K]$ containing
counterterms intended to cancel loop infinities. Both $S_R$ and
$S_\infty$ must have the same symmetries as $S[\chi,K]$, so the
infinite contributions to $\Gamma$ can be canceled by the
counterterms in $S_\infty$ if they also have those symmetries. 

Expanding $\Gamma$ in a power series in the loop expansion
parameter $\hbar$,
\begin{eqnarray}
\Gamma[\chi, K] = \sum_{N=0}^\infty \hbar^{N-1} \Gamma_N[\chi, K]\,,
\end{eqnarray}
where $\Gamma_0[\chi, K] = S_R[\chi, K]$, the Zinn-Justin equation
can be written order-by order for each $N$ as 
\begin{eqnarray}
\sum_{N'=0}^N \left(\Gamma_{N'}, \Gamma_{N - N'}\right) = 0 \,.
\label{sym.sumn'}
\end{eqnarray}
This expansion automatically includes counterterms corresponding to
sub-divergences at any given loop order $N$. If for some $N$ all
infinities appearing at $M$-loop order have been canceled by
counterterms in $S_\infty$ for all $M \leq N-1$, the only remaining
infinities in Eqn.(\ref{sym.sumn'}) are in $\Gamma_N$. So the
infinite part $\Gamma_{N,\infty}$ of this quantity must satisfy 
\begin{eqnarray}
\left( S_R, \Gamma_{N,\infty} \right) = 0\,.
\label{sym.infinite}
\end{eqnarray}

For a theory which is renormalizable in the power-counting sense,
this leads to a simple mechanical procedure. For such a theory,
the infinite part $\Gamma_{N,\infty}[\chi, K]$ must be a sum of
operators of mass dimension four or less. In addition, all the linear
symmetries of the tree-level action are symmetries of $\Gamma[\chi,
K]$ and therefore of $\Gamma_{N,\infty}[\chi, K]$.

Let me assume for the moment that  $\Gamma_{N,\infty}[\chi, K]$ is
at most linear in the antisources $K^A$ for all $A$, 
\begin{eqnarray}
\Gamma_{N,\infty}[\chi, K] = \Gamma_{N,\infty}[\chi, 0] + \int d^4x
{\mathscr F}^A_N[\chi, x] K^A(x) \,.
\end{eqnarray}
If I now define the quantities 
\begin{eqnarray}
\Gamma_N^{(\epsilon)}[\chi] = S_R[\chi, 0] + \epsilon
\Gamma_{N,\infty}[\chi, 0] \,,
\label{sym.gammaep}
\end{eqnarray}
with $\epsilon$ infinitesimal, the terms independent of $K^A$ in
Eqn.(\ref{sym.infinite}) imply\cite{weinqft} that
$\Gamma_N^{(\epsilon)}[\chi]$ is invariant under the transformation 
\begin{eqnarray}
s_R \chi^A(x) = F_N^{(\epsilon)A}(x) \,,
\end{eqnarray}
where
\begin{eqnarray}
F_N^{(\epsilon)A}(x) = F^A(x) + \epsilon{\mathscr F}^A_N(x) \,.
\label{sym.Fep}
\end{eqnarray}
The terms of first order in $K^A$ in Eqn.(\ref{sym.infinite}) imply
that this transformation is nilpotent, $s_R^2 = 0$.
Since $\Gamma_{N,\infty}$ contains only operators of mass dimension
four or less, $F_N^{(\epsilon)A}(x)$ cannot be of higher mass
dimension than $F^A(x)$. In addition, $F_N^{(\epsilon)A}(x)$ may
not affect the linear symmetries of the action. Therefore,
$F_N^{(\epsilon)A}(x)$ must have the same Lorentz properties, ghost
number and global gauge transformation properties as $F^A(x)$. In
fact $F_N^{(\epsilon)A}(x)$ must be the same as $F^A(x)$ if it
corresponds to a field which transforms linearly under $s$. All
that remains to be done is to construct the most general nilpotent
transformation of the fields under these restrictions, and then to
construct the most general functional $\Gamma_N^{(\epsilon)}[\chi]$
invariant under this transformation. If that agrees, up to
arbitrary constant numerical coefficients, with the original action
$S$, the theory is perturbatively renormalizable.

This entire argument rests on the assumption that $\Gamma_{N,
\infty}[\chi, K]$ is at most linear in all of the antisources
$K^A$. When is this a correct assumption? If a field $\chi^A$ has
mass dimension $d_A$, the corresponding $K^A$ must have mass
dimension $3 - d_A$ so as to make $\int d^4 x F^A K^A$
dimensionless. The antisources $K^A$ for $ A^a_\mu, \omega^a,
\bw^a, B^a_{\mu\nu}, \omega^a_\mu, \bw^a_\mu, n^a, \beta^a, \bb^a $
all have mass dimension $2$. The antisources for $C^a_\mu$ and
$\theta^a$ each have mass dimension 3.  Also, the theory does not have
any external antisource $K^A$ for the fields $h^a, h^a_\mu, \alpha^a,
\ba^a$ because their BRST variations vanish. Therefore $\Gamma_{N,
\infty}$ can be at most quadratic in $K^A$.

If a field $\chi^A$ has ghost number $\gamma_A$, the corresponding
$K^A$ will have ghost number $-\gamma_A -1$. It follows that the
ghost number of the antisource for any of $A^a_\mu, B^a_{\mu\nu},
C^a_\mu, n^a$ is $-1$. The ghost numbers of $K^A$ corresponding to
$\omega^a, \omega^a_\mu$ and $\theta^a$ is $-2$, and those of $K^A$
corresponding to $\beta^a$ and $\bb^a$ are $-3$ and $+1$,
respectively.  The remaining antisources correspond to $\bw^a$ and
$\bw^a_\mu$, they carry ghost number 0. The dimensions and ghost
numbers of all the fields and their antisources are given in Table
\ref{a3.t.mdgn} at the end of this paper.

Some of the quadratic terms can be eliminated straightaway.  The
BRST variations of the fields $\bw^a, \bw^a_\mu, n^a$ and $\bb^a$
are linear, so the effective action cannot be quadratic in their
antisources. For example,
\begin{eqnarray}
s\bw^a = -h^a, 
\end{eqnarray}
so the quantum transformations are the same,
\begin{eqnarray}
\langle s\bw^a \rangle_{J_{\chi,K},K} = - h^a \,.
\end{eqnarray}
It follows from Eqn.(\ref{sym.neweffac})that 
\begin{eqnarray}
{\delta_R \Gamma[\chi,K] \over \delta K^a[\bw] } = -h^a \,,
\end{eqnarray}
for the corresponding antisource $K^a[\bw]$. Since this independent
of $K^A$, it follows that $\Gamma[\chi,K]$ is linear in the
antisource for $\bw^a$. A similar argument holds for $\bw^a_\mu,
n^a$ and $\bb^a$. Let me now look at the antisources for the
remaining fields in the theory. The quantum effective action
$\Gamma[\chi, K]$ must be linear in the antisources of $\theta^a$
and $C^a_\mu$, since these objects have mass dimension 3 and all
other antisources have mass dimension 2. So $\Gamma[\chi, K]$ is at
most quadratic in the antisources of only the other fields. It
turns out that $\Gamma[\chi, K]$ is in fact linear in the remaining
antisources as well. The argument involves showing that the
coefficients of the quadratic terms are forced to vanish, term by
term, by the dimensions and ghost numbers of the fields which can
possibly appear in them.  Appendix \ref{a1}\ contains the details
of the argument.

It follows then that the effective action is at most linear in all
the antisources $K^A$, and the arguments
following~Eqn.(\ref{sym.infinite}) hold. But the number of possible
terms in the effective action allowed by the (renormalized) BRST
symmetry $s_R$ is still enormous, and it is necessary to invoke
other symmetries to simplify calculations.

Let me now consider the effect of the gauge-dependent symmetry
$\ts$ on the effective action. I take the same partition function
$Z[J,K]$ and the same effective action $\Gamma[\chi, K]$ as in
Eqs.~(\ref{sym.newZ})\ and (\ref{sym.neweffac}), with the same
sources $ J^A_{\chi,K}$ and the same antisources $K^A$. (This
$\Gamma[\chi, K]$ was shown to be linear in these $K^A$ in
Appendix~\ref{a1}.) Let me also denote the minimal fields by
$\phi^A$ and non-minimal fields by $\lambda^A$. Then $\ts\phi^A =
s\phi^A = F^A$, and consequently $\ts F^A[\phi] = 0$. The
application of $\ts$ on the partition function gives (since the
tree-level action $S$ is invariant under $\ts$),
\begin{eqnarray}
- \int d^4x \left[\langle F^A \rangle_{J_{\chi,K},K}
{\delta_L\Gamma[\chi, K]\over \delta \phi^A}+ \langle \ts\lambda^A
\rangle_{J_{\chi,K},K} {\delta_L\Gamma[\chi, K]\over \delta
\lambda^A} + \langle \ts s\lambda^A \rangle_{J_{\chi,K},K}
K^A[\lambda]\right] = 0\,.
\label{sym.ts}
\end{eqnarray} 
Now, if the gauge-fixing functions are linear in the fields,
$\ts\lambda^A$ as defined in Eqn.(\ref{brs.stilde}) is either linear
in the fields or equals the BRST variation of some linear function of
the fields. Therefore, $\langle \ts \lambda^A \rangle_{J_{\chi, K},
K}$ is known in principle from solving the Zinn-Justin equations. In
addition, the effective action does not contain the antisources
corresponding to $(h^a, h^a_\mu, \alpha^a, \ba^a)$ and only $S_R$
contains the antisources for $(\bw^a, \bw^a_\mu, n^a,\bb^a)$. Then I
can read off from Eqn.(\ref{sym.ts})\ that $\Gamma^{(\epsilon)}_N
[\chi]$ as defined in Eqn.(\ref{sym.gammaep})\ is invariant under
$\ts_R$, which is just $\ts$ as calculated in terms of $s_R$. In other
words, $\Gamma^{(\epsilon)}_N$ is invariant under $\ts_R$ where
\begin{eqnarray}
\ts_R \bw^a &=&  h^a + {2\over \xi} \, f^a \,, \nonumber \\ 
\ts_R h^a &=& -{2\over \xi} s_Rf^a \equiv -{2\over \xi} \Delta^a_R
\,, \nonumber \\ 
\ts_R \bw^a_\mu &=& h^a_\mu + {2\over \eta} \, \partial_\mu n^a + 
{2\over \eta} \, f^a_\mu \,, \nonumber \\ 
\ts_R h^a_\mu &=& {2\over \eta} \left( \partial_\mu \alpha^a +
{2\over \zeta} \, \partial_\mu f'^a - s_Rf^a_\mu \right) \,
\equiv  {2\over \eta} \left( \partial_\mu \alpha^a +
{2\over \zeta} \, \partial_\mu f'^a - \Delta^a_{R\mu} \right) \,,
\nonumber \\ 
\ts_R n^a &=& - \left( \alpha^a + {2\over \zeta} \, f'^a \right)
\,, \nonumber \\ 
\ts_R \bb^a &=& - \left( \ba^a - {2\over \zeta} \,,
\partial_\mu \bw^{a\mu} \right) \nonumber \\ 
\ts_R \alpha^a &=& - {2\over \zeta} s_Rf'^a \equiv  - {2\over
\zeta} \Delta'^a_R \,, \nonumber \\ 
\ts_R \ba^a &=& {2\over \zeta}\, \partial_\mu \left( h^{a\mu}
+ {2\over \eta } \, \partial^\mu n^a + {2\over \eta} f^{a\mu}
\right) \,, 
\label{sym.stilde} \\
\ts_R &=& s_R \mbox{ on all other fields.} \nonumber
\end{eqnarray}
Note that I did not fully utilize the nilpotence of $\ts$ itself. In
principle, I could have treated $\ts$ just like $s$, defining
new antisources $\widetilde K^A$ and deriving an analogue of
Zinn-Justin equation. But that creates a host of other problems. In
particular, the effective action is not linear in these new
antisources $\widetilde K^A$.

These two renormalized symmetries, $s_R$ and $\ts_R$ are sufficient
to uniquely fix the form of the effective action, as will be shown
in the next section.  There is a further symmetry which helps to
pin down the form of $s_R$. This symmetry mixes the ghost fields
with the same global properties and quantum numbers.

The action $S$ is invariant under 
\begin{eqnarray}
\delta\bw^a &=& -\dl\,\ba^a \,,\nonumber \\
\delta\omega^a_\mu &=& \dl\,\left(\partial_\mu\omega^a + g
f^{abc}A^b_\mu\omega^c \right) \,, \nonumber \\
\delta\theta^a &=& -\dl\,\omega^a \,, \nonumber \\
\delta\beta^a &=& \dl\,\half g f^{abc}\omega^b\omega^c \,, 
\label{sym.mixing} \\
\delta{\rm (all\, others)} &=& 0 \,,\nonumber 
\end{eqnarray}
where $\dl$ is a {\em commuting} c-number infinitesimal.
It is straightforward to calculate that 
\begin{eqnarray}
ts\theta^a = \half gf^{abc}\omega^b\omega^c, \qquad ts {\rm (all\,
others)} = 0 \,,
\label{sym.tschi}
\end{eqnarray}
where $t$ is the transformation $\delta/\dl$. Note that I have
taken $\dl$ to be commuting only for convenience. If $\dl$ is taken
to be anticommuting, the action will still be symmetric under $t =
\delta_L/\dl$ provided $\delta_L \bw^a/\dl = + \ba^a$, other
transformation rules remaining the same.  It is easy to see that
the action $S$ is symmetric under $t$ for a large class of
gauge-fixing functions $f^{a\mu}$.

By applying $t$ on the partition function (\ref{sym.newZ}), I get
the Ward identities
\begin{eqnarray}
\int d^4x \left( \langle t\bw^a \rangle\, {\delta_L\Gamma\over
\delta\bw^a} + \langle t\omega^a_\mu \rangle {\delta_L\Gamma\over
\delta\omega^a_\mu} + \langle t\theta^a \rangle\,
{\delta_L\Gamma\over \delta\theta^a} + \langle t\beta^a \rangle\,
{\delta_L\Gamma\over \delta\beta^a} - \langle ts \theta^a \rangle\,
K^a[\theta]
\right)  = 0 \,,
\end{eqnarray}
where the quantum averages $\langle\,\rangle$ are calculated in the
presence of the currents and antisources $J_{\chi, K}, K$ as
before, and $ K^a[\theta]$ is the antisource for
$\theta^a$. Since $ t\theta^a = -\omega^a$, $t\bw^a = -\ba^a$ are
linear in the fields, their quantum averages are the same. As for
the other two, $ t\omega^a_\mu = sA^a_\mu$ and $t\beta^a = -
s\omega^a$, so the quantum averages of the quantities on the left
hand side are known. I can then write this equation as
\begin{eqnarray}
\int d^4x\,\left(  - \ba^a {\delta_L\Gamma\over \delta\bw^a}
+ {\delta_R\Gamma \over \delta K^{a\mu}[A]}
{\delta_L\Gamma \over \delta\omega^a_\mu}
-\omega^a  {\delta_L\Gamma\over \delta\theta^a}
- {\delta_R\Gamma \over \delta K^a[\omega]}  {\delta_L\Gamma\over
\delta\beta^a}
+ {\delta_R\Gamma \over \delta K^a[\omega]}  K^a[\theta] \right)
= 0 \,.
\end{eqnarray}

Expanding $\Gamma$ in a power series in $\hbar$ and using arguments
as before, I can write the divergent part of this equation as
\begin{eqnarray}
\int d^4x \left( -  \ba^a {\delta_L\Gamma_{N, \infty}\over
\delta\bw^a}
+ {\delta_R S_R \over \delta K^{a\mu}[A]}
{\delta_L\Gamma_{N, \infty} \over \delta\omega^a_\mu} \right.
&+& {\delta_R\Gamma_{N, \infty} \over \delta K^{a\mu}[A]}
{\delta_L S_R \over \delta\omega^a_\mu}
- \omega^a  {\delta_L\Gamma_{N, \infty}\over \delta\theta^a}
\,\nonumber \\
-  {\delta_R S_R \over \delta K^a[\omega]}
{\delta_L\Gamma_{N,\infty}\over \delta\beta^a}
&-&  {\delta_R\Gamma_{N, \infty} \over \delta K^a[\omega]}  {\delta_L
S_R \over \delta\beta^a} 
+\left. {\delta_R\Gamma_{N, \infty} \over \delta K^a[\omega]}
K^a[\theta] \right) = 0 \,.
\end{eqnarray}
The $K$ independent terms of this equation give 
\begin{eqnarray}
\int d^4x \left(- \ba^a {\delta_L\Gamma^{(\epsilon)}[\chi]\over
\delta\bw^a} - F^{(\epsilon)a}_\mu[A]
{\delta_L\Gamma^{(\epsilon)}[\chi] \over \delta\omega^a_\mu}  
- \omega^a  {\delta_L\Gamma^{(\epsilon)}[\chi]\over \delta\theta^a} 
- F^{(\epsilon)a}[\omega]{\delta_L\Gamma^{(\epsilon)}[\chi]\over
\delta\beta^a} \right) = 0 \,,
\label{sym.zeroth}
\end{eqnarray}
where $\Gamma^{(\epsilon)}[\chi]$ and $ F^{(\epsilon)}[\chi]$ are
as defined in Eqs.~(\ref{sym.gammaep}) and (\ref{sym.Fep}) with the
indices $N$ and $x$ suppressed, and I have used the invariance of
$S_R$ under the transformation $t$. The terms of first order in the
antisources give
\begin{eqnarray}
\int d^4x \left( - \ba^a {\delta_L\over
\delta\bw^a} - F^{(\epsilon)a}_\mu[A]
{\delta_L \over \delta\omega^a_\mu}  
- \omega^a {\delta_L\over \delta\theta^a} 
- F^{(\epsilon)a}[\omega]{\delta_L\over
\delta\beta^a} \right) F^{(\epsilon)B}[\chi] =0 \,
\label{sym.first1}
\end{eqnarray}
for all $B$ except when $B$ corresponds to $\theta^a$, where I have
used the fact that $ts = 0$ on all fields except $\theta^a$. For
the case of $K^a[\theta]$, this equation is modified,
\begin{eqnarray}
\int d^4x \left\{ \left(- \ba^b {\delta_L\over
\delta\bw^b} - F^{(\epsilon)b}_\mu[A]
{\delta_L \over \delta\omega^b_\mu}  
- \omega^b {\delta_L\over \delta\theta^b} 
- F^{(\epsilon)b}[\omega]{\delta_L\over
\delta\beta^b} \right) F^{(\epsilon)a}[\theta]
+  F^{(\epsilon)a}[\omega] \right\} =0 \,,
\label{sym.first2}
\end{eqnarray}
upon using $ts\theta^a = \half g f^{abc}\omega^b\omega^c \equiv
-F^a[\omega]$. 

The interpretation of these equations is obvious. 
Eqn.(\ref{sym.zeroth}) says that $\Gamma^{(\epsilon)}_N[\chi]$ is
invariant under $t_R$, where
\begin{eqnarray}
t_R\bw^a &=& -\ba^a \,,\nonumber \\
t_R\omega^a_\mu &=& s_R A^a_\mu \,, \nonumber \\
t_R\theta^a &=& - \omega^a \,, \nonumber \\
t_R\beta^a &=& - s_R \omega^a \,, 
\label{sym.tR} \\
t_R{\rm (all\, others)} &=& 0 \,,\nonumber 
\end{eqnarray}
The first equation following that, Eqn.(\ref{sym.first1}), shows
that $t_Rs_R = 0$ on all fields except $\theta^a$, and
Eqn.(\ref{sym.first2}) shows that $t_Rs_R\theta^a =
-s_R\omega^a$. There are no surprises, except perhaps the fact that
these conditions are actually useful in restricting the form of
$s_R$ to what is shown in Appendix~\ref{a2}.

\section{The Most General Effective Action } \label{ren}
In this section I shall use brute force methods to show that the
effective action contains the same terms, up to arbitrary
multiplicative renormalizations, as the tree-level action of
Eqn.(\ref{brs.treeS}). The proof requires construction of the most
general nilpotent transformation of the fields, $\sr$, as discussed
in the previous section. The actual construction of $\sr$ is a
rather involved digression, so I have separated it into
Appendix~\ref{a2}. The result of $\sr$ on the various fields is
given in Eqn.(\ref{a2.sr}).

Now I need to construct the most general functional
$\Gamma_N^{(\epsilon)}$ symmetric under $\sr$ as well as under the
linear symmetries of $S$. These are $(i)$ Lorentz invariance, $(ii)$
global $SU(N)$ invariance, and $(iii)$ ghost number
conservation. However, it is obvious that even with the restrictions
imposed by $\sr$ and these three symmetries, there is an enormous
number of possible terms. Since $C^a_\mu$ has mass dimension zero as
well as ghost number zero, it is possible to multiply any other term
by a scalar polynomial of the form $\sum\limits_{k=1}^n (C^a_\mu
C^{a\mu})$, and still maintain the three abovementioned symmetries. Of
course, such terms are not $\sr$-invariant by themselves, but one can
imagine that their variations cancel against those of something else,
and ruling out each such term requires a long and tedious
calculation. And even without this unwanted complexity, there are of
the order of one hundred terms satisfying the three linear
symmetries. Applying $\sr$ to a sum of so many terms, multiplied by
unknown scalar polynomials of $C^a_\mu$, and then finding the
combination which remains invariant, would require unlimited time and
perseverance. Fortunately, there is a way out of this quagmire,
provided by $\ts$.

As before, let me denote the minimal fields by $\phi^A$ and
non-minimal fields by $\lambda^A$. Let me also define $\sr' \equiv
\half(\sr - \tilde\sr)$. Then from Eqn.(\ref{sym.ts}),
\begin{eqnarray}
\sr'\phi^A = 0\,, \sr'\lambda^A \neq 0\,.
\label{ren.diffsym}
\end{eqnarray}
The non-minimal fields $\lambda^A$ are $(\bw^a, \bb^a, \bw^a_\mu,
\ba^a, h^a, \alpha^a, h^a_\mu, n^a)$. Let me also choose the gauge
fixing functions to be specifically those in
Eqn.~(\ref{brs.gauges}). Because of my choice of gauge-fixing
functions, the action exhibits invariance under constant
shifts of $\bw^a, \bb^a, \bw^a_\mu$ and $n^a$. Since these are
linear symmetries, I can impose them on the effective action. In
other words, these fields must appear in the quantum effective
action only as derivatives, i.e., as $\partial_\mu\bw^a$ etc.

Then on dimensional grounds, the effective action will be at most
quadratic in the $\lambda^A$. I can then write the effective action
in the generic form
\begin{eqnarray}
\Gamma = \sum_A \lambda^A X^A + \sum_{A,B}\lambda^A\lambda^B X^{AB}\,,
\label{ren.effac}
\end{eqnarray}
where $X^A$ and $X^{AB}$ do not contain any of the $\lambda^A$, and
have appropriate transformation properties, dimension and ghost
number. In particular, $X^A$ and $X^{AB}$ are assumed to include
derivative operators as necessary for the constant shift symmetries
mentioned above, and the sum over indices will be taken to include
an integral over space-time unless specified otherwise. Since both
$\sr$ and $\tilde\sr$ are symmetries of the effective action, I
have
\begin{eqnarray}
\sr'\Gamma = 0 \,,
\end{eqnarray}
and from Eqn.(\ref{ren.diffsym}), I have
\begin{eqnarray}
\sr'X^A = \sr' X^{AB} = 0 \,.
\label{ren.srallX}
\end{eqnarray}
Therefore using Eqn.(\ref{ren.effac}) I can write
\begin{eqnarray}
\sum\limits_A (\sr'\lambda^A)X^A + \sum\limits_{A,B} (\sr'\lambda^A)
\lambda^B X^{AB} +  \sum\limits_{A,B} (-1)^{\varepsilon_A} \lambda^A
(\sr'\lambda^B) X^{AB} = 0\,.
\label{ren.expand}
\end{eqnarray} 
Here $\varepsilon_A$ (not to be confused with the $\varepsilon_A$
of Eqs.~(\ref{sym.firstord}) and (\ref{sym.quadord})) is the
Grassmann parity of the field $\lambda^A$. 

Since $X^A$ and $ X^{AB}$ do not contain any of the $\lambda^A$ by
definition, I can now look at the coefficients of the various
$\lambda^A$ in the expansion of Eqn.(\ref{ren.expand}) and set them
to zero in order to get an expression for the effective action
$\Gamma$. Many of the terms are thus eliminated, and some algebraic
relations appear among some of the rest. The details of the
calculation are given in Appendix \ref{a3}. The result is the
effective Lagrangian of Eqn.(\ref{a3.ghostac}) which I give here
again,
\begin{eqnarray}
{\mathscr L}_g = \bw^a X^a_{\bw} &+& \bb^a X^a_{\bb} + \bw^a_\mu
X^{a\mu}_{\bw_*} + \ba^a X^a_{\ba} + h^a X^a_h + h^a_\mu
X^{a\mu}_{h_*} + \partial_\mu n^a X^{a\mu}_n \, \nonumber \\
&+&  \bw^a_\mu \alpha^b X^{ab\mu}_{\bw_*\alpha} + \ba^a \alpha^b
X^{ab}_{\ba\alpha} + h^a h^b X^{ab}_{hh} + h^a \left( \eta h^b_\mu
+ \partial_\mu n^b \right) X^{ab\mu}_{hn} \,\nonumber \\
&+& h^a_\mu h^b_\nu
X^{ab\mu\nu}_{h_* h_*} + h^a_\mu\partial_\nu n^b X^{ab\mu\nu}_{h_*
n} + \partial_\mu n^a \partial_\nu n^b X^{ab\mu\nu}_{nn} \,.
\label{SR'GAMMA}
\end{eqnarray} 
The undetermined coefficients satisfy several relations among
themselves as shown in Appendix~\ref{a3}. 

The number of unknown coefficients can be reduced even further.
Just as the symmetry $\sr'\Gamma = 0$ produced relations among
several of these $X$'s, the quantum BRST symmetry itself,
$\sr\Gamma = 0$, should produce some more relations independent of
the previous ones. The expression for the $\sr$-variation is
\begin{eqnarray}
\sr\Gamma = && \sum\limits_A^{}\left[ (\sr\lambda^A)X^A +
(-1)^{\varepsilon_A} \lambda^A(\sr X^A) \right] \,\nonumber \\
&&+ \sum\limits_{AB}^{}
\left[ (\sr\lambda^A)\lambda^B X^{AB} + (-1)^{\varepsilon_A}\lambda^A
(\sr\lambda^B) X^{AB} + (-1)^{\varepsilon_A + \varepsilon_B}\lambda^A
\lambda^B (\sr X^{AB}) \right] = 0.
\end{eqnarray} 
Since $\sr X^A$ and $\sr X^{AB}$ do not contain any of the
$\lambda^A$, I can consider the coefficients of $\lambda^A$ or of
$\lambda^A\lambda^B$ in the above expression and set them to zero.

The calculation is fairly straightforward, but in keeping with
other calculations in this paper, I have again separated this one
into Appendix~\ref{a4}. The result is that the functional forms of
all the unknown coefficients become known, and only two arbitrary
constants are needed to write them, as shown in
Eqn.~(\ref{a3.lgfinal}).

I can now write down the general form of the ghost sector of the
theory as 
\begin{eqnarray}
{\mathscr L}_g = Z_\omega\, \bw^a \Delta^a_R &+& 
Z_\beta\, \bb^a \Delta'^a_R + Z_\beta\, \bw^a_\mu
\Delta^{a\mu}_R + Z_\beta\, \ba^a f'^a + Z_\omega\,
h^a f^a + Z_\beta\, h^a_\mu \left( f^{a\mu} + \partial^\mu
n^a \right) \nonumber \\
&-& Z_\beta\, \partial^\mu\bw^a_\mu \alpha^a + \zeta\, 
Z_\beta\, \ba^a\alpha^a + {\xi\over 2}\,Z_\omega\, h^ah^a  
+ {\eta\over 2}\,Z_\beta\, h^a_\mu h^{a\mu}\,.
\label{LGFINAL}
\end{eqnarray} 
It remains to construct the most general non-ghost sector of the
theory. 

The BRST transformations on the bosonic fields as found in
Eqn.(\ref{a2.sr}) of the Appendix are given by 
\begin{eqnarray}
\sr A^a_\mu &=& \partial_\mu\omega^a_R + g_R
f^{abc} A^b_\mu\omega^c_R \,,\nonumber \\
\sr B^a_{\mu\nu} &=& g_R f^{abc} B^b_{\mu\nu}\omega^c_R +
\partial_{[\mu}\omega^a_{R\nu]} + g_R f^{abc}
A^b_{[\mu}\omega^c_{R\nu]} 
+ g_R f^{abc}\partial_{[\mu}A^b_{\nu]}
\theta^c_R + g_R^2 f^{aed}f^{ebc}A^b_\mu A^c_\nu \theta^d_R ) 
\,,\nonumber \\ 
\sr C^a_\mu &=& g_R f^{abc}C^b_\mu \omega^c_R + 
{{\mathscr N}_4\over {\mathscr N}_6}\,\left(\omega^a_{R\mu} +
\partial_\mu\theta^a_R + g_R f^{abc}A_\mu^b\theta^c_R \right)\,,
\end{eqnarray} 
where I have defined $g_R = \displaystyle{g\over{\mathscr N}_1},\,
\omega^a_R = {\mathscr Z}{\mathscr N}_1\omega^a,\, \omega^a_{R\mu}
= {\mathscr Z}{\mathscr N}_6 \omega^a_\mu$ and $\theta^a_R =
{\mathscr Z}{\mathscr N}_6{\mathscr N}_1 \theta^a$. If I now define
{\em renormalized} field strengths
\begin{eqnarray}
\tilde F^a_{\mu\nu} &=& \partial_\mu A_\nu^a - \partial_\nu A_\mu^a +
g_R f^{abc}A_\mu^b A_\nu^c\, \nonumber \\
\tilde H_{\mu\nu\lambda}^a &=& \partial_{[\mu}B_{\nu\lambda]}^a +
g_R f^{abc}A^b_{[\mu}B^c_{\nu\lambda]}
- {{\mathscr N}_6\over {\mathscr N}_4}\, g_R
f^{abc}\tilde F_{[\mu\nu}^bC_{\lambda]}^c \,, 
\end{eqnarray} 
I find that the Zinn-Justin equation just says that the ghost-free
sector is invariant under these gauge transformations. The factor
${\mathscr N}_6/{\mathscr N}_4$ can be absorbed either in $C^a_\mu$ itself or
in the renormalization of a fiduciary coupling constant $g_C$ which
always appears in front of $C^a_\mu$.

The procedure described so far can be used to construct effective
actions for different theories involving the non-Abelian
two-form. For example, it may be interesting to apply it to the
recently proposed first-order formulation of Yang-Mills theory
\cite{catzen}. However, since I have a specific theory in mind, I
will need to invoke another symmetry in order to eliminate unwanted
terms from the non-ghost sector.

This `symmetry' was an invariance of the classical equations of
motion under 
\begin{equation}
B^a_{\mu\nu} \to B^a_{\mu\nu} + \alpha F^a_{\mu\nu}
\label{ren.shift}
\end{equation}
with $\alpha$ a constant. It was suggested in \cite{gvm}\ that this
symmetry could play a role in preventing terms  of the form $(B -
DC)^2$ or $(B - DC)\wedge(B - DC)$ from appearing in the action. Of
course, since the classical action is not invariant but changes by
a total derivative, it is nontrivial to elevate this to a quantum
symmetry. Classically this `symmetry' leads to a conserved current
\begin{equation}
J^\mu_T = H^{a\mu\nu\lambda}F^a_{\nu\lambda} +
m\epsilon^{\mu\nu\lambda\rho} \big(A^a_\nu F^a_{\lambda\rho} -
{2\over 3}f^{abc}A^a_\nu A^b_\lambda A^c_\rho \big) \,.
\end{equation}
But there is no corresponding conserved current for the BRST-invariant 
action $S$ of Eqn.(\ref{brs.treeS}), which was the starting point of
quantization. This part of the problem can be circumvented quite
easily by incorporating the shift into the BRST transformation for
$B^a_{\mu\nu}$, which now reads (cf. Eqn.(\ref{brs.brst}))
\begin{equation}
sB^a_{\mu\nu} = gf^{abc}B^b_{\mu\nu}\omega^c +
\partial_{[\mu}\omega^a_{\nu]}  + gf^{abc}A^b_{[\mu}\omega^c_{\nu]} +
gf^{abc}F^b_{\mu\nu}\theta^c + \alpha F^a_{\mu\nu}.
\label{ren.Bbrs}
\end{equation}
Here $\alpha$ is an anticommuting {\em constant} with ghost number
+1, and $s\alpha = 0$. It is trivial to see that the BRST transformation
$s$ is still nilpotent, $s^2 = 0$. However, the action is still not
invariant, but is shifted by ${{m\alpha\over 4}}\int d^4x
\epsilon^{\mu\nu\lambda\rho}F^a_{\mu\nu}F^a_{\lambda\rho}$. So it
seems that I have not gained anything, but only recovered a symmetry
of the equations of motion. On the other hand, since I am now dealing
with the quantum theory rather than the classical action principle, I
can also generate this term through quantum effects. 

For example, this term could be canceled by the transformation of
the fermion measure if fermions are coupled to the gauge field.
Under a chiral transformation with a parameter $4\pi^2m\alpha$, the
effective action changes by $- {m\alpha\over 4}\int
d^4x\epsilon^{\mu\nu\lambda\rho} F^a_{\mu\nu}F^a_{\lambda\rho}$,
which cancels the effect of the shift transformation of
Eqn.~(\ref{ren.shift}). The action then becomes invariant under the
combination of the shift and global chiral transformations. There
are other ways of canceling the term generated by the shift. In any
case, symmetry under the shift transformation rules out terms
involving products of $\big(B_{\mu\nu} - D_{[\mu}C_{\nu]}\big)$. I
can then write down the most general quantum effective action
consistent with the quantum symmetries,
\begin{equation}
\Gamma_{\mathit eff} = \int d^4x\, \Big( Z_A \tilde F^a_{\mu\nu}\tilde
F^{a\mu\nu} +  Z_B \tilde H^a_{\mu\nu\lambda} \tilde
H^{a\mu\nu\lambda} + Z_{BF} m\epsilon^{\mu\nu\lambda\rho}
B^a_{\mu\nu} \tilde F^a_{\lambda\rho} + {\mathscr L}_g \Big)\,.
\end{equation}
This is the same as the tree-level action up to arbitrary
multiplicative constants, which means that the theory is
perturbatively renormalizable.

\section{Discussion of results}\label{disc}
\hyphenation{re-norma-liz-abi-lity}

It is time to gather the results. I have given an algebraic proof
of perturbative renormalizability of the dynamical non-Abelian
two-form gauge theory, also known as the topological mass
generation mechanism. It follows that just as in two and three
dimensions, it is possible in four dimensions to have a
renormalizable theory of massive non-Abelian vector bosons without
a residual Higgs particle.

The calculations were done in a specific set of linear gauges, so
that antighosts appeared only as derivatives.  In other linear
gauges, the calculations would be more involved, in particular
there would be terms cubic and quartic in the antighosts, but even
in such cases the methods of Sec.~\ref{ren}\, should go through.
Two other symmetries appeared as a result of using linear gauges ---
these are $\ts$, defined in Eqn.~(\ref{brs.stilde}), and $t$,
defined in Eqn.~(\ref{sym.mixing}).  These two symmetries were
greatly useful for constructing the quantum BRST symmetry and for
reducing the number of possible terms in the quantum effective
action. These symmetries would be present in other linear gauges as
well, but not in a general nonlinear gauge. The calculations are
extremely tedious for nonlinear gauges, and it is not clear if the
quantum BRST symmetry alone is sufficient to restrict the terms in
the quantum effective action to the same form as those in the
tree-level action in such gauges.

The proof also depends crucially on the nature of the auxiliary
field $C^a_\mu$. There is no quadratic term for this field in the
action of Eqn.~(\ref{brs.S0}), and it was mentioned in
Sec.~\ref{lag} that as a result there was no propagator for
$C^a_\mu$ and diagrams with internal $C^a_\mu$ lines vanished. This
seems rather peculiar, but it is actually not a problem for
perturbation theory as long as there are other fields which
propagate freely. The free Hamiltonian can always be written in
terms of the propagating free fields, and other terms can be
thought of as perturbation on top of it, with $C^a_\mu$ being a
non-dynamical field.  One may question whether perturbation theory
is valid for an action for which the operator in the matrix of
quadratic terms is not invertible, as in this case. In general,
perturbation can be done only if a free Hamiltonian can be
constructed for the theory. I have assumed that this is a
sufficient condition as well, as long as all the {\em physical}
fields appear in the free Hamiltonian and their quadratic matrix
can be inverted. The free Hamiltonian is a sum of terms like
$\half(\Pi^2 + \Phi^2)$ only over the {\em physical} degrees of
freedom, and this is the part which gives the propagators. So if
the number of degrees of freedom in the theory is known, it is
strictly necessary to have only that many propagators, and
therefore only that many quadratic pieces in the theory. The
quadratic matrix of the physical fields in this theory can be
inverted, as can be seen from the fact that perturbatively there
are three propagating degrees of freedom, as there should be from
counting constraints. In other words, the assumption made here is
that for a perturbation series to be constructed, only physical
degrees of freedom need to be identified with quantum fields. The
non-propagating degrees need not be quantum fields in the sense of
canonical quantization as long as the path integral can be formally
constructed for them. This is not a radical assumption, it is made
for all gauge theories, but is usually associated with unphysical
objects (such as the scalar mode of a vector field) which are not
Lorentz covariant.
On the other hand, it is clear that $C^a_\mu$ by itself is not a
physical field, as it can be completely removed by a vector gauge
transformation. 

Therefore I can try to choose a gauge in which the path integral
over $C^a_\mu$ can be formally calculated in the Lagrangian
formalism and in which there is a propagator for $C^a_\mu$.  The
gauge chosen in Sec.~\ref{brs} was $f^a_\mu = \partial^\nu
B^a_{\mu\nu}$ which did not give a propagator for $C^a_\mu$. For
convenience, let me keep to linear gauges so that the proof given
above can be used with minimal modifications. Since $C^a_\mu$ can
be shifted away, it would seem that the gauge $C^a_\mu = 0$,
i.e. the choice of $f^a_\mu = m^2C^a_\mu$ could provide the
necessary term. However, this is not a good gauge choice. The
vector gauge transformations get fixed completely, but now there is
no propagator for $B^a_{\mu\nu}$ as the corresponding quadratic
operator is non-invertible. One possible alternative is to choose
an $R_\xi$ type gauge, $f^a_\mu = \partial^\nu B^a_{\mu\nu} + \eta
m^2 C^a_\mu$. In this gauge there is no canonical momentum for
$C^a_\mu$ (which was true for the classical theory also), so
$C^a_\mu$ terms do not appear in the free Hamiltonian (unlike the
apparently analogous case of a Goldstone mode in broken gauge
theories). But formally the Lagrangian path integral over $C^a_\mu$
can be done, and formally $C^a_\mu$ will have a propagator as well
as a two-point vertex with $B^a_{\mu\nu}$ after using the equations
of motion for $h^a_\mu$ and $n^a$. The total tree-level propagator
will be a sum over insertions of these vertices as well as vertices
between $A^a_\mu$ and $B^a_{\mu\nu}$. This propagator will be
gauge-dependent,
\begin{eqnarray}
\widetilde\Delta^{ab}_{C\mu\nu} = - {\delta^{ab}\over m^4}{k^2 -
m^2 \over (\eta -1)k^2 - \eta m^2}\,.
\label{disc.Rxim}
\end{eqnarray}
This propagator clearly does not represent a real particle because
it has a gauge dependent pole, as is expected because of the
Goldstone-like nature of $C^a_\mu$. This propagator also has the
ultraviolet behavior of $\Delta^{ab}_{C\mu\nu} \sim O(1)$ as $k^\mu
\to \infty$, which is not very good for proofs of
renormalizability, although because of the gauge-dependence of the
pole in $\Delta^{ab}_{C\mu\nu}$ it may not be relevant. However, in
this gauge the action is not invariant under a constant shift of
$\bw^a_\mu$, so the proof given above will not be applicable
directly but will have to be redone after including cubic and
quadratic terms in $\bw^a_\mu$.

Another possibility for an $R_\xi$ type gauge choice is $f^a_\mu =
\partial^\nu B^a_{\mu\nu} + \eta \Box C^a_\mu$. The gauge-fixed
theory in this gauge also does not have a canonical momentum for
$C^a_\mu$ and therefore terms involving $C^a_\mu$ do not appear in
the free Hamiltonian. But again the Lagrangian path integral over
$C^a_\mu$ can be done formally, and after summing over all
two-point vertex insertions, the total tree-level propagator will
again be gauge-dependent,
\begin{eqnarray}
\Delta^{ab}_{C\mu\nu} = - {\delta^{ab}\over k^4}{k^2 - m^2 \over
(\eta -1)k^2 - \eta m^2}\,.
\label{disc.Rxik}
\end{eqnarray}
This (formal) propagator also has a gauge dependent pole and
therefore cannot represent a real particle. On the other hand,
$\Delta^{ab}_{C\mu\nu}$ falls off as $O(k^{-4})$ in the ultraviolet
regime. Since $C^a_\mu$ has been assumed to have a vanishing mass
dimension, $\Delta^{ab}_{C\mu\nu}$ therefore satisfies the
criterion for power-counting renormalizability that given a field
of mass dimension $d$, its propagator should fall off as
$O(k^{-\delta})$, where $4 - \delta \leq 2d$~\cite{piso}.
Therefore it can be safely used in algebraic proofs of
renormalizability. And because of this particular choice of gauge,
the antighost $\bw^a_\mu$ still appears only with derivatives, and
the action is still symmetric under a constant shift of
$\bw^a_\mu$, and the effective action is still at most quadratic in
$\bw^a_\mu$. Because this is a linear gauge, the rest of the
arguments of the paper go through without modification, and we
recover the same proof in this $R_\xi$-type gauge. In both these
gauges, the propagators for the ghosts of the vector symmetry also
have gauge-dependent poles.

Although the Lagrangian path integral over $C^a_\mu$ can be
formally done in the above gauges, the nature and the role of the
auxiliary field remain obscure. Even then this is somewhat better
than a canonical analysis of the system, where it is not possible
to identify $C^a_\mu$ with a quantum field because its canonical
momentum vanishes. That quantization is problematic even in the
Lagrangian path integral formalism is brought out by the fact that
the calculations above remain the same even in the absence of a
propagator for $C^a_\mu$. However, the essential point of this
paper is the following --- despite apparent problems with
perturbative expansion of this theory, application of standard
algebraic techniques to it leads to a quantum action which contains
the same operators as the classical action, which would be a proof
of renormalizability in any other theory. This result was
completely unexpected on the grounds of the problems with
perturbative expansion, already mentioned above and in~\cite{nogo}.

The use of the anticommuting constant $\alpha$ can potentially
create problems because it has vanishing mass dimension, but ghost
number +1, by Eqn.~(\ref{ren.Bbrs}). The fact that there is an
anticommuting constant in the theory is not in itself a problem,
similar objects appear in supersymmetric quantum
mechanics~\cite{nickm}. But the fact that it has vanishing mass
dimension can cause problems of its own. One place where problems
can arise is the argument in Sec.~\ref{sym} and particularly in
Appendix~\ref{a1} that the quantum effective action is at most
linear in the antisources $K^A$. The argument relied on the fact
that the coefficients of these quadratic terms had vanishing
mass dimension but non-vanishing ghost number, so they must contain
$\theta^a$. Now $\theta^a$ can be replaced by the constant
$\alpha$. But there is no reason to worry, because the relevant
objects have ghost number +2 or more, so at least one $\theta^a$
will be needed to construct any of them, and the rest of the
argument remains unchanged.  Another possible place for a problem
is in the calculation of the general nilpotent transformation
$\sr$, given in the Appendix~\ref{a2}. Some of the fields $\chi^A$
could have a term like $\alpha\chi^A$ in their transformation rules
in principle. Other similarly constructed terms are also possible.
An explicit calculation using the $t$ symmetry shows that such
terms do not arise.

I have not touched on the issue of anomalies, or the inclusion of
fermions, in detail. Fermions will couple to the Yang-Mills gauge
field in the usual way, but there is no gauge-invariant coupling of
mass dimension four between the two-form and fermions because of
the shift symmetry mentioned in Eqn.~(\ref{ren.shift}). The
Yang-Mills theory will have the usual SU(N) anomaly of $Tr~F\wedge
F$. This can be removed by use of the shift symmetry. Gauge
anomalies will be absent if the gauge group is the Standard Model
gauge group. The two-form brings with it a vector gauge symmetry,
as given in Eqn.~(\ref{brs.vxfn}). This is an Abelian symmetry, but
there is no field carrying the charge corresponding to it. So there
is no anomaly involving this transformation. 

The original motivation for the theory was to find a possible
alternative for the Higgs sector of the Salam-Weinberg model of
electroweak interactions. What I have shown in this paper is that
it is possible to have massive vector bosons without spontaneous
symmetry breaking. But the problem with applying this mechanism to
electroweak interactions is precisely that there is no symmetry
breaking, whereas the observed world has broken SU(2)$\times$U(1)
symmetry. If I write the Lagrangian for the Salam-Weinberg model
without the Higgs field, and add an SU(2) two-form with an action
as in Eqn.(\ref{brs.S0}), all the SU(2) gauge bosons get the same
mass, contrary to experiment. All other {\em observed} events would
remain uncontradicted. It has been suggested~\cite{neto}\ that by
also adding a U(1) two-form, which would make the photon massive,
it may be possible to get the correct mass ratio of Z and W$^\pm$
particles. However, agreement with experiment requires an infinite
parameter in the classical Lagrangian, corresponding to an infinite
mass for the photon. Another way is to add an explicit symmetry
breaking term to the Lagrangian, so that the mass term reads
\begin{equation}
{m \over 4}\epsilon^{\mu\nu\rho\lambda}
\left(B^a_{\mu\nu}F^a_{\rho\lambda} -  \tan\theta_{_W} 
B^3_{\mu\nu} F^{}_{\rho\lambda}\right)\,,
\label{disc.symmbreak}
\end{equation}
where $F_{\rho\lambda}$ is the field strength of the U(1) gauge
field, and $\theta_{_W}$ is the Weinberg angle. Then by defining
the Z and photon fields as usual, I can write the quadratic part of
this mass term as
\begin{equation}
{\epsilon^{\mu\nu\rho\lambda}\over 4}\left( m B^1_{\mu\nu}
\partial_{[\rho}A^1_{\lambda]} + m B^2_{\mu\nu}\partial_{[\rho}
A^2_{\lambda]} + m \sec\theta_{_W}B^3_{\mu\nu}
\partial_{[\rho}Z_{\lambda]} \right)\,.
\end{equation} 
It follows from this that the Z is heavier than the W$^\pm$ by a
factor of $\sec\theta_{_W}$ (and the photon is massless). However,
because of the explicit symmetry breaking term, the proof of
renormalizability given here is not applicable. So the question of
applicability of this model to electroweak interactions remains
open.

{\em Acknowledgement:} It is a pleasure to acknowledge long
discussions with F.~Barbero and E.~Sanchez about the nature of the
auxiliary vector field.


\newpage
\appendix

\section{Antisource dependence of $\Gamma_{N,\infty}$}\label{a1}

As was argued in Sec.~\ref{sym}, the quantum effective action is at
most quadratic in the antisources. In fact, several of the
quadratic terms were eliminated just by looking at the mass
dimensions and BRST transformation properties of the fields. In
order to see the dependence of the effective action on the rest of
the antisources, let me write the general expression of $\Gamma_{N,
\infty}[\chi, K]$  as
\begin{eqnarray}
\Gamma_{N, \infty}[\chi, K] = \Gamma_{N, \infty}[\chi, 0] +  \int d^4x\,
{\mathscr F}^A_N[\chi, x] K^A(x) +  \int d^4x\,
{\mathscr F}^{AB}_N[\chi, x] K^A(x)K^B(x) \,. 
\label{sym.quadgamma}
\end{eqnarray}
In this there is no $K^A$ corresponding to $h^a, h^a_\mu, \alpha^a$
and $\ba^a$, and the quadratic sum also does not run over the
antisources for $\theta^a, C^a_\mu, \bw^a, \bw^a_\mu, n^a$ and
$\bb^a$ for reasons described in Sec.~\ref{sym}.  The relation $\left(S_R,
\Gamma_{N,\infty}\right) = 0$, when applied to this expression,
gives at zeroth order in $K^A$
\begin{eqnarray}
- \int d^4x\,{\mathscr F}_N^A\, {\delta_L S_R[\chi, 0]\over \delta
\chi^A} - \int d^4x\,F^A\, {\delta_L \Gamma_{N, \infty}[\chi, 0] 
\over \delta \chi^A} = 0 \,.
\label{sym.qinv}
\end{eqnarray}
At first order in $K^A$, I get the equation 
\begin{eqnarray}
\int d^4x \left[{\mathscr F}_N^B\, {\delta_L F^A \over \delta \chi^B}
+ F^B \, {\delta_L {\mathscr F}_N^A\over \delta \chi^B} + 2{\delta_L
S_R[\chi, 0]\over \delta \chi^B}\, {\mathscr F}_N^{AB} \right] = 0 \,.
\label{sym.firstord}
\end{eqnarray}
Here I have used the fact that $ {\mathscr F}_N^{BA} =
(-1)^{\varepsilon_A\varepsilon_B}\,{\mathscr F}_N^{AB}$ where
$\varepsilon_A, \varepsilon_B$ are the Grassmann parities of $K^A$
and $K^B$, $0$ for bosonic $K^A$ and $1$ for fermionic $K^A$.  The
terms of second order in the antisources lead to the equation
\begin{eqnarray}
\sum_C\int d^4x\,F^C(x)\,{\delta_L {\mathscr
F}_N^{AB}(y)\over \delta  
\chi^C(x)}&&\delta^4(y -z) \nonumber\\
+ \sum_C &&\left[(-1)^{\varepsilon_A( \varepsilon_B + \varepsilon_C
+ 1)} {\mathscr F}_N^{AC}(y)\,{\delta_L F^B(z) \over \delta \chi^C(y)}
+ A\leftrightarrow B \right] = 0 \,.
\label{sym.quadord}
\end{eqnarray}
The coefficient ${\mathscr F}^{AB}_N[\chi, x]$ has mass dimension $d_A
+ d_B - 2$, and ghost number $\gamma_A + \gamma_B + 2$, where $d_A$
and $\gamma_A$ are respectively the mass dimension and ghost number
of the field $\chi^A$.  Since $\theta^a$ has ghost number $+1$ and
mass dimension zero, it is possible to construct functions of
arbitrary positive ghost number and mass dimension zero by taking
products of $\theta^a$.  Since the quadratic sum runs only over the
antisources for the fields ${A^a_\mu, B^a_{\mu\nu}, \omega^a,
\omega^a_\mu, \beta^a}$, it follows that ${\mathscr F}_N^{AB}$ can
depend {\em only} on $\theta^a$ and $C^a_\mu$ for all $A,B$, and
$\theta^a$ must be present in ${\mathscr F}_N^{AB}$ to take care of its
ghost number, which is always positive. So the first term of
Eqn.~(\ref{sym.quadord}), $F^C \, \delta_L {\mathscr F}_N^{AB}/ \delta
\chi^C$, must contain for all $A,B$,
\begin{eqnarray}
(s\theta^a)\, {\delta_L {\mathscr F}_N^{AB}\over \delta \theta^a}
= ( - gf^{abc}\theta^b\omega^c - \beta^a)  {\delta_L {\mathscr
F}_N^{AB}\over \delta \theta^a} \,.
\end{eqnarray} 
The first term on the right hand side will always appear in $ {\mathscr
F}^{AC}\, {\delta_L F^B / \delta \chi^C}$ because $ {\mathscr F}^{AC}$
contains $\theta^a$ for all $A,B$, but the second will appear only if
$F^A$ contains $\beta^a$. (The index $N$ is suppressed from now on.)

In the sum $ {\mathscr F}^{AC}\, {\delta_L F^B /\delta \chi^C}$, the
only terms that contribute a $\beta^a$ are for $\chi^C$
corresponding to $A^a_\mu$ when $\chi^B$ is $\omega^a_\mu$, and
$\chi^C$ corresponding to $\omega^a$ when $\chi^B$ is
$\beta^a$. This implies, first of all, that at least one of the
indices $A,B$ in ${\mathscr F}^{AB}$ must correspond to either
$\omega^a_\mu$ or $\beta^a$. In other words, when neither $\chi^A,
\chi^B$ corresponds to $\omega^a_\mu$ or $\beta^a$, the sum ${\mathscr
F}^{AC}\, {\delta_L F^B / \delta\chi^C}$ does not contain $\beta^a$
even after (anti-)symmetrization over $A,B$, while the sum $F^C \,
\delta_L {\mathscr F}^{AB}/ \delta \chi^C$ must contain
$\beta^a$. Therefore $ {\mathscr F}^{AB} = 0$ for all such pairs $A,B$.
So for example  $ {\mathscr F}^{ab}(\omega,\omega) = 0$.

Now, the only ${\mathscr F}^{AC}$ which contribute a $\beta^a$ to the sum
${\mathscr F}^{AC}\, {\delta_L F^B /\delta \chi^C}$ are those for which
one index corresponds to one of $(\omega^a_\mu,\beta^a)$, and the
other index to $A^a_\mu$ or $\omega^a$ and all these ${\mathscr F}^{AC}$
contain only products of $\theta^a$ and $C^a_\mu$.  Looking at ${\mathscr
F}^{AB}$ of this type, I find that each term which can contribute a
$\beta^a$ to the sum has a factor ${\mathscr F}^{AC}$ of the type that
vanishes by the previous argument. For example, if $(A,B)$ correspond
to $(\omega^a, \beta^a)$, the only term in the sum that could
contribute a factor of $\beta^a$ is ${\mathscr F}^{ac}(\omega,
\omega)\, {\delta_L F^b(\beta)/\delta \omega^c}$, which vanishes
since ${\mathscr F}^{ac}(\omega, \omega) = 0$. Explicitly, for this
case Eqn.(\ref{sym.quadord}) reads
\begin{eqnarray}
\int {\mathscr F}^{ac}(\omega, \omega)\, {\delta_L F^b(\beta) \over
\delta \omega^c} + \int {\mathscr F}^{ac}(\omega, \beta)\, {\delta_L
F^b(\beta) \over \delta \beta^c} &+& \int {\mathscr F}^{ac}(\omega,
\beta)\, {\delta_L  F^b(\omega) \over \delta \omega^c} \,\nonumber
\\
&+& \int F^c(\chi){\delta_L {\mathscr F}^{ab}(\omega, \beta)\over
\delta\chi^c}  = 0\,.
\label{sym.example}
\end{eqnarray}
The first term vanishes because ${\mathscr F}^{ac}(\omega, \omega) =
0$ by the previous argument, the second and the third terms cannot
contain a $\beta^a$, while the fourth term must contain only one
factor of  $\beta^a$. Since this is impossible, ${\mathscr
F}^{ab}(\omega, \beta)$ must also vanish. It follows by a similar
argument that  ${\mathscr F}^{AB} = 0$ when both indices correspond
to  $\omega^a_\mu$ or $\beta^a$.

So the sum ${\mathscr F}^{AC}\, {\delta_L F^B /\delta \chi^C}$ vanishes
for all $(A,B)$, and therefore ${\mathscr F}^{AB} = 0$ for all pairs
$(A,B)$. It follows that  $\Gamma_{N,\infty}[\chi, K]$, and hence
the quantum effective action, is at most linear in the antisources
$K^A$, so that the arguments following Eqn.~(\ref{sym.infinite})
can be used towards a proof of renormalizability.

\section{Renormalized BRST transformation} \label{a2}
I need to construct a generalized BRST transformation of the
fields. This is a nilpotent transformation which affects the
Lorentz properties, ghost numbers and global gauge transformation
properties of the fields in exactly the same way as $s$ of
Eqn.(\ref{brs.brst}) and is identical with the latter where it is
linear. Let me calculate the generalized nilpotent transformation
$\sr$ for one field at a time. For the fields which transform
linearly under BRST, this is the same as the original $s$,
\begin{eqnarray}
\sr\bw^a = -h^a\,, \qquad \sr h^a = 0\,, \qquad \sr\bw^a_\mu =
-h^a_\mu \,, \qquad \sr h^a_\mu = 0\,, \nonumber \\
\sr n^a = \alpha^a\,, \qquad \sr\alpha^a = 0\,,\qquad \sr\bb^a =
\ba^a \,, \qquad \sr\ba^a = 0\,.
\end{eqnarray}

For the gauge field $A_\mu^a$ and the associated ghost $\omega^a$,
I can write
\begin{eqnarray}
\sr A_\mu^a &=& b_1^{ab} \partial_\mu\omega^b + g
d_1^{abc}A_\mu^b\omega^c \,,\nonumber \\
\sr \omega^a &=& -\half g d_2^{abc}\omega^b\omega^c \,.
\label{a2.Aomega}
\end{eqnarray}
The nilpotence condition $\sr^2 \omega^a = 0$ implies
\begin{eqnarray}
d_2^{abc}d_2^{cde} + d_2^{adc}d_2^{ceb} + d_2^{aec}d_2^{cbd} = 0
\,. 
\label{a2.bianchi}
\end{eqnarray}
Therefore $d_2^{abc}$ must be proportional to the structure
constants $f^{abc}$,
\begin{eqnarray}
d_2^{abc} = {\mathscr Z}\,f^{abc}\,,
\label{a2.d2}
\end{eqnarray}
where ${\mathscr Z}$\ is an arbitrary constant. In $\sr^2 A_\mu^a = 0$,
the coefficient of $A_\mu^d \omega^e \omega^c$ gives
\begin{eqnarray}
d_1^{abc} d_1^{bde} - d_1^{abe} d_1^{bdc} = {\mathscr Z}\, d_1^{adb}
f^{bec}\,, 
\label{a2.firstd}
\end{eqnarray}
which has the unique solution
\begin{eqnarray}
d_1^{abc} = {\mathscr Z}\, f^{abc},
\label{a2.d1}
\end{eqnarray}
while the coefficient of $\partial_\mu\omega^c\omega^d$ gives
\begin{eqnarray}
b_1^{ab}f^{bcd} = f^{abd}b_1^{bc}\,,
\label{a2.firstb}
\end{eqnarray}
which implies
\begin{eqnarray}
b_1^{ab} = {\mathscr Z}{\mathscr N}_1\,\delta^{ab}\,,
\label{a2.b1}
\end{eqnarray}
with ${\mathscr N}_1$\, again an arbitrary constant. 

Let me now write the rules for $\beta^a$ and $\theta^a$, 
\begin{eqnarray}
\sr \beta^a &=& g d_3^{abc} \beta^b \omega^c \,,\nonumber \\
\sr \theta^a &=& - g d_4^{abc} \theta^b \omega^c - b_2^{ab}\beta^b \,.
\label{a2.betatheta}
\end{eqnarray}
In $\sr^2 \beta^a = 0$, the coefficient of $\beta^d\omega^e
\omega^c$ gives as in Eqn.(\ref{a2.firstd}),
\begin{eqnarray}
d_3^{abc} = {\mathscr Z}\, f^{abc}\,.
\label{a2.d3}
\end{eqnarray}
In $\sr^2 \theta^a = 0$, the coefficient of $\theta^d \omega^e
\omega^c$ gives as in Eqn.(\ref{a2.firstd}),
\begin{eqnarray}
d_4^{abc} =  {\mathscr Z}\, f^{abc}\,,
\label{a2.d4}
\end{eqnarray}
while the coefficient of $\beta^c \omega^d$ gives as in
Eqn.(\ref{a2.firstb}),
\begin{eqnarray}
b_2^{ab} = {\mathscr Z}{\mathscr N}_2 \delta^{ab}\,.
\label{a2.b2}
\end{eqnarray}
For the fields $\omega_\mu^a$ and $C_\mu^a$ the rules are
\begin{eqnarray}
\sr \omega^a_\mu &=& - g d_5^{abc} \omega_\mu^a \omega^c +
b_3^{ab}\partial_\mu\beta^b + g d_6^{abc} A_\mu^b \beta^c
\,,\nonumber \\
\sr C^a_\mu &=& g d_7^{abc} C^b_\mu \omega^c + b_4^{ab}\omega_\mu^b +
b_5^{ab}\partial_\mu\theta^b + g d_8^{abc} A_\mu^b\theta^c\,.
\label{a2.Comega}
\end{eqnarray}
In $\sr^2 \omega^a_\mu = 0$, the coefficient of $\omega^d_\mu
\omega^e \omega^c$ gives as in Eqn.(\ref{a2.firstd}),
\begin{eqnarray}
d_5^{abc} = {\mathscr Z}\, f^{abc}\,,
\label{a2.d5}
\end{eqnarray}
the coefficient of $\partial_\mu\beta^c \omega^d$ gives as in
Eqn.(\ref{a2.firstb}) 
\begin{eqnarray}
b_3^{ab} = {\mathscr Z}{\mathscr N}_3\, \delta^{ab}\,,
\label{a2.b3}
\end{eqnarray}
with ${\mathscr N}_3$ arbitrary, the coefficient of $\beta^d \partial_\mu 
\omega^e$ gives upon using Eqn.(\ref{a2.b3}),
\begin{eqnarray}
d_6^{abc} = {{\mathscr Z}{\mathscr N}_3 \over {\mathscr N}_1}\, f^{abc}\,,
\label{a2.d6}
\end{eqnarray}
and the coefficient of $A_\mu^d \beta^e \omega^c$ vanishes
identically as a result.

In $\sr^2 C^a_\mu = 0$, the coefficient of $C^d_\mu \omega^e
\omega^c$ gives as in Eqn.(\ref{a2.firstd}),
\begin{eqnarray}
d_7^{abc} = {\mathscr Z}\, f^{abc}\,,
\label{a2.d7}
\end{eqnarray}
the coefficients of $\omega^d_\mu \omega^c$ and
$\partial_\mu\theta^d \omega^c$ give as in Eqn.(\ref{a2.firstb}),
\begin{eqnarray}
b_4^{ab} = {\mathscr Z}{\mathscr N}_4\, \delta^{ab}\,, \nonumber \\
b_5^{ab} = {\mathscr Z}{\mathscr N}_5\, \delta^{ab},  
\label{a2.b4b5}
\end{eqnarray}
where ${\mathscr N}_4$ and ${\mathscr N}_5$ are arbitrary
constants. The coefficient of $\partial_\mu\beta^b$ then gives 
\begin{eqnarray}
{\mathscr N}_3{\mathscr N}_4 = {\mathscr N}_2{\mathscr N}_5\,,
\label{a2.NN}
\end{eqnarray}
the coefficient of $A_\mu^b\beta^c$ gives 
\begin{eqnarray}
d_8^{abc} = {{\mathscr Z}{\mathscr N}_5\over {\mathscr N}_1}\,f^{abc}\,,
\label{a2.d8}
\end{eqnarray}
and the coefficients of $A_\mu^d\theta^e\omega^c$ and $\theta^b
\partial_\mu \omega^c$ vanish identically as a result.

Finally, $B^a_{\mu\nu}$ transforms as
\begin{eqnarray}
\sr B^a_{\mu\nu} = g d_9^{abc}B^b_{\mu\nu}\omega^c &+& b_6^{ab} 
\partial_{[\mu}\omega^b_{\nu]} + g d_{10}^{abc} A^b_{[\mu}
\omega^c_{\nu]} \, \nonumber \\
&+& g d_{11}^{abc}\partial_{[\mu}A^b_{\nu]}\theta^c 
+g^2 e^{abcd}A^b_\mu A^c_\nu \theta^d \,.
\label{a2.Bmn} 
\end{eqnarray}
The constant $ e^{abcd}$ is antisymmetric in two indices, $e^{acbd}
= - e^{abcd}$. The coefficient of $B^d_{\mu\nu}\omega^e\omega^c$ in
$\sr^2 B^a_{\mu\nu} = 0$ gives as in Eqn.(\ref{a2.firstd})
\begin{eqnarray}
d_9^{abc} = {\mathscr Z}\, f^{abc}\,,
\label{a2.d9}
\end{eqnarray}
the coefficient of $\partial_{[\mu}\omega^d_{\nu]}\omega^c$ gives
as in Eqn.(\ref{a2.firstb}),
\begin{eqnarray}
b_6^{ab} = {\mathscr Z}{\mathscr N}_6\,\delta^{ab}\,,
\label{a2.b6}
\end{eqnarray}
the coefficient of $(\partial_{[\mu}\omega^d)\omega^c_{\nu]}$ gives
upon using Eqn.(\ref{a2.b6})
\begin{eqnarray}
d_{10}^{abc} = {{\mathscr Z}{\mathscr N}_6\over {\mathscr N}_1}\,f^{abc}
\,, 
\label{a2.d10}
\end{eqnarray}
the coefficient of $\partial_{[\mu}A_{\nu]}^c\beta^d$ gives upon
using Eqn.(\ref{a2.b6})
\begin{eqnarray}
d_{11}^{abc} = {{\mathscr Z}{\mathscr N}_3{\mathscr N}_6\over {\mathscr
N}_1{\mathscr N}_2}\, f^{abc}\,,
\label{a2.d11}
\end{eqnarray}
the coefficient of $A_\mu^b \partial_\nu \omega^c \theta^d$ gives
upon using Eqn.(\ref{a2.d11})
\begin{eqnarray}
e^{abcd} = {{\mathscr Z}{\mathscr N}_3{\mathscr N}_6\over {\mathscr
N}_1^2{\mathscr N}_2}\, f^{aed}f^{ebc}\,.
\label{a2.eabcd}
\end{eqnarray}
All other coefficients in the expression of $\sr B^a_{\mu\nu}$
vanish identically as a result.

Note that it is possible to consider other terms in $s_R$ which
obey the Zinn-Justin equation at first order in the antisources. I
have ignored such terms because they vanish upon using the symmetry
$t$. Let me consider one example, that of $\omega^a$. The
Zinn-Justin equation says that 
\begin{eqnarray}
s{\mathscr F}^a[\omega] - gf^{abc}{\mathscr F}^b[\omega]\omega^c = 0 \,.
\end{eqnarray}
On the other hand, from Eqn.(\ref{sym.first1}) for $\omega^a$, I
have 
\begin{eqnarray}
t{\mathscr F}^a[\omega] = 0\,.
\end{eqnarray}
The only allowed possibility for ${\mathscr F}^a[\omega]$ is then 
\begin{eqnarray}
{\mathscr F}^a[\omega] = -{1 \over 2\epsilon} g (d_2^{abc} - f^{abc}) 
\omega^b \omega^c  + e^{abc}(\omega_\mu + D_\mu\theta)^b
(\omega^\mu + D^\mu\theta)^c \,,
\end{eqnarray}
where $d_2^{abc}$ and $e^{abc}$ are now arbitrary.
Eqn.(\ref{sym.first2}) gives 
\begin{eqnarray}
t{\mathscr F}^a[\theta] + 2 {\mathscr F}^a[\omega] = 0\,,
\end{eqnarray}
which immediately shows that $ e^{abc} = 0$, and I can write
Eqn.~(\ref{a2.Aomega})\, for the transformation of $\omega^a$.

Another byproduct of this equation is the somewhat unexpected
relation 
\begin{eqnarray}
{\mathscr Z}\,{\mathscr N}_2 = 1 \,.
\end{eqnarray}
Similarly, I can use Eqn.(\ref{sym.first1}) to relate some of the
constants previously found. {}From $t_Rs_R \omega^a_\mu = 0$, I find
${\mathscr N}_1 = {\mathscr N}_3$, and from $t_Rs_R C^a_\mu = 0$, I
find ${\mathscr N}_5 = {\mathscr Z}\,{\mathscr N}_1{\mathscr N}_4$. No other
new relation can be found this way.

The transformation rules can now be collected,
\begin{eqnarray}
\sr A^a_\mu &=& {\mathscr Z}\,({\mathscr N}_1\partial_\mu\omega^a + g
f^{abc} A^b_\mu\omega^c) \,,\nonumber \\
\sr \omega^a &=& -\half {\mathscr Z}\,g f^{abc}\omega^b \omega^c \,,
\qquad \sr\bw^a = -h^a \,, \qquad \sr h^a = 0 \,, \nonumber \\
\sr B^a_{\mu\nu} &=& {\mathscr Z}\,( g f^{abc} B^b_{\mu\nu}\omega^c +
{\mathscr N}_6 \partial_{[\mu}\omega^a_{\nu]} +  {{\mathscr N}_6 \over 
{\mathscr N}_1}\, g f^{abc} A^b_{[\mu}\omega^c_{\nu]} \,\nonumber \\
&&\qquad + {\mathscr Z}\,{\mathscr N}_6\, 
g f^{abc}\partial_{[\mu}A^b_{\nu]}
\theta^c + {{\mathscr Z}\,{\mathscr N}_6\over{\mathscr N}_1}\,
 g^2 f^{aed}f^{ebc}A^b_\mu
A^c_\nu \theta^d )\,,\nonumber \\ 
\sr C^a_\mu &=& {\mathscr Z}\,(g f^{abc}C^b_\mu \omega^c + 
{\mathscr N}_4\,\omega^a_\mu + {\mathscr Z}\,{\mathscr N}_1\,{\mathscr
N}_4\, \partial_\mu\theta^a +
{{\mathscr Z}\, {\mathscr N}_4}\, g f^{abc}A_\mu^b\theta^c )\,,
\nonumber \\
\sr \omega^a_\mu &=& {\mathscr Z}\,(-g f^{abc}\omega^b_\mu \omega^c 
+ {\mathscr N}_1\, \partial_\mu\beta^a + 
g f^{abc} A^b_\mu\beta^c) \,, \nonumber\\
\sr \bw^a_\mu &=& - h^a_\mu \,, \qquad \sr  h^a_\mu = 0 \,, \qquad
\sr n^a = \alpha^a \,, \qquad \sr \alpha^a = 0 \,, \nonumber \\
\sr \beta^a &=&  {\mathscr Z}\, g f^{abc}\beta^b \omega^c \,, \qquad
\sr \bb^a = \ba^a \,, \qquad \sr \ba^a = 0\,, \nonumber \\
\sr \theta^a &=&  - {\mathscr Z}\,g f^{abc}\theta^b \omega^c - \beta^a \,. 
\label{a2.sr}
\end{eqnarray}
%

\section{Derivation of Eqn.(\ref{sr'Gamma})}\label{a3}
The generic form of the effective action is
\begin{eqnarray}
\Gamma = \sum_A \lambda^A X^A + \sum_{A,B}\lambda^A\lambda^B X^{AB}\,,
\label{a3.effac}
\end{eqnarray}
where $X^A$ and $X^{AB}$ do not contain any of the $\lambda^A$.
Therefore as mentioned in Sec.\ref{ren}\ I can write the effect of
$\sr'$ on $\Gamma$ as 
\begin{eqnarray}
\sum\limits_A (\sr'\lambda^A)X^A + \sum\limits_{A,B} (\sr'\lambda^A)
\lambda^B X^{AB} +  \sum\limits_{A,B} (-1)^{\varepsilon_A} \lambda^A
(\sr'\lambda^B) X^{AB} = 0\,.
\label{a3.expand}
\end{eqnarray} 
Since $X^A$ and $X^{AB}$ do not contain any of the $\lambda^A$, I
can look at the coefficients of the various $\lambda^A$ in the
expansion of Eqn.(\ref{a3.expand}) and set them to zero in order to
get an expression for the effective action $\Gamma$. The effect of
$\sr'$ on $\lambda^A$ is, for quick reference,
\begin{eqnarray}
\sr'\bw^a &=& -\left( h^a + {1\over\xi}f^a \right)\,,
\qquad \sr'\bb^a = \left( \ba^a -
{1\over\zeta}\partial_\mu\bw^{a\mu} \right) \,, \nonumber\\
\sr'\bw^a_\mu &=& -\left(h^a_\mu + {1\over \eta}\partial_\mu n^a +
{1\over \eta}f^a_\mu \right)\,, \qquad
\sr'\ba^a = -{1\over \zeta} \partial^\mu \left(h^a_\mu + {2\over
\eta}\partial_\mu n^a + {2\over \eta}f^a_\mu \right)\,, 
\nonumber \\
\sr' h^a &=& {1\over \xi}\Delta^a_R\,, \qquad \sr'\alpha^a = {1\over
\zeta} \Delta'^a_R \,, \nonumber \\
\sr' h^a_\mu &=& - {1\over \eta} 
\left( \partial_\mu \alpha^a + {2\over \zeta}\partial_\mu f'^a -
\Delta^a_{R\mu} \right) \,, \qquad \sr'n^a = \left(\alpha^a +
{1\over \zeta} f'^a \right)\,.
\end{eqnarray} 

In this Appendix, I will construct the most general $\Gamma$
obeying Eqn.(\ref{a3.expand}). For each $\lambda^A$ I will first
consider coefficients of terms containing $\lambda^A\lambda^B$ in
the expansion of Eqn.(\ref{a3.expand}).  There can be no term of
third or higher order in $\lambda^A$ in the effective action
because of the constant shift symmetries, and therefore the left
hand side of Eqn.(\ref{a3.expand}) can be at most quadratic in the
$\lambda^A$.  The coefficients will have as many ghost fields and
derivative operators as necessary.  Setting the coefficients to
zero will eliminate some of the terms from the effective action and
produce relations among some others.  Following the same procedure
for the terms linear in $\lambda^A$ will produce some more
relations. There will also be a few terms not containing any power
of $\lambda^A$ in the expansion of Eqn.(\ref{a3.expand}). The sum
of these should also vanish.

I will consider $\lambda^A$ in the order $(\bw^a, \bb^a, \bw^a_\mu,
\ba^a, h^a, \alpha^a, h^a_\mu, \partial_\mu n^a)$. Terms containing
products of the form $\bw^a\lambda^B$ in the expansion of
$\sr'\Gamma$ come from the $\sr'$-variation of
\begin{eqnarray}
\bw^a\bw^b X^{ab}_{\bw\bw} + \bw^a\bb^b X^{ab}_{\bw\bb} +
\bw^a\ba^b X^{ab}_{\bw\ba} + \bw^a\bw^b_\mu X^{ab\mu}_{\bw\bw_*} + 
\bw^a h^b_\mu X^{ab\mu}_{\bw h_*} + \bw^a\partial_\mu n^b
X^{ab\mu}_{\bw n}\,, 
\label{a3.bwquad}
\end{eqnarray} 
where the subscripts $\bw\bw$ etc. indicate the quadratic
combination which couples to a given $X$, and an asterisk indicates
the presence of a Lorentz index on the subscript.  In the first
term, $ X^{ab}_{\bw\bw}$ has to be antisymmetric in $[a,b]$.
Therefore, the coefficient of $\bw^ah^b$ in the expansion of
$\sr'\Gamma$ gives $X^{ab}_{\bw\bw}=0$. The coefficients of
$\bw^a\ba^b, \bw^a h^b_\mu$ and $\bw^a\partial_\mu n^b$ give
\begin{eqnarray}
X^{ab}_{\bw\bb} =  X^{ab}_{\bw\ba} =  X^{ab\mu}_{\bw\bw_*} = 0\,,
\end{eqnarray} 
while the coefficient of $\bw^a\alpha^b$ gives  $X^{ab\mu}_{\bw
h_*} = \eta X^{ab\mu}_{\bw n}$. Terms linear in $\bw^a$ (and
containing no other $\lambda^A$) come from 
\begin{eqnarray}
\bw^a\alpha^b X^{ab}_{\bw\alpha} + \bw^a h^b X^{ab}_{\bw h} + 
(\eta\bw^a h^b_\mu + \bw^a \partial_\mu n^b) X^{ab\mu}_{\bw n}
\label{a3.bwlin}
\end{eqnarray} 
The coefficient of $\bw^a$ in $\sr'\Gamma$ is therefore
\begin{eqnarray}
-{1\over\zeta} \Delta'^b_R  X^{ab}_{\bw\alpha} - {1\over \xi}
\Delta^b_R  X^{ab}_{\bw h} + \left( {1\over\zeta} \partial_\mu f'^b
- \Delta^b_{R\mu} \right) X^{ab\mu}_{\bw n} = 0.
\end{eqnarray} 
Each of the $X$'s in this equation must contain at least one
derivative operator to allow for the constant shift symmetry of
$\bw^a$. Therefore, the $X$'s must be constructed only out of
fields of mass dimension zero. Since the factors multiplying the
$X$'s are all different, and have fields of non-vanishing mass
dimension, the only choice for which this equation can be satisfied
is
\begin{eqnarray}
 X^{ab}_{\bw\alpha} =  X^{ab}_{\bw h} =  X^{ab\mu}_{\bw n} = 
X^{ab\mu}_{\bw h_*} = 0\,.
\end{eqnarray} 
Thus, all terms containing $\bw^a\lambda^B$ are excluded from the
effective action.

Terms containing $\bb^a\lambda^B$ arise from the terms 
\begin{eqnarray}
\bb^a\bb^b X^{ab}_{\bb\bb} + \bb^a\bw^b_\mu X^{ab}_{\bb\bw_*} 
+ \bb^a\ba^b X^{ab}_{\bb\ba} + \bb^a h^b_\mu X^{ab\mu}_{\bb h_*}
+ \bb^a\partial_\mu n^b X^{ab\mu}_{\bb n}
\label{a3.bbquad}
\end{eqnarray} 
As before, I set the coefficients of the quadratic terms in the
expansion  to zero. The coefficients of $\bb^a\ba^b, \bb^a h^b_\mu$
and $\bb^a \partial_\mu n^b$ give 
\begin{eqnarray}
X^{ab}_{\bb\bb} =  X^{ab}_{\bb\bw_*} =  X^{ab}_{\bb\ba} = 0 \,,
\end{eqnarray} 
while the coefficient of $\bb^a\ba^b$ gives $ X^{ab\mu}_{\bb h_*} =
\eta  X^{ab\mu}_{\bb n}$.

Terms linear in $\bb^a$ appear from the $\sr'$-variation of 
\begin{eqnarray}
\bb^a h^b X^{ab}_{\bb h} + \bb^a\alpha^b X^{ab}_{\bb\alpha} - 
\left( \eta\bb^a h^b_\mu + \bb^a \partial_\mu n^b \right)
X^{ab\mu}_{\bb n} \,.
\label{a3.bblin}
\end{eqnarray} 
The coefficient of $\bb^a$ in the variation of this is 
\begin{eqnarray}
{1\over\xi}\Delta^b_R X^{ab}_{\bb h} + {1\over\zeta} \Delta'^b_R
X^{ab}_{\bb\alpha} - \left( {1\over \zeta}\partial_\mu f'^b -
\Delta^b_{R\mu} \right) X^{ab\mu}_{\bb n} = 0 \,.
\end{eqnarray} 
Again each of the $X$'s in this equation has mass dimension zero
after excluding the derivative operator they must contain to allow
a constant shift in $\bb^a$. It is easy to see that the only
solution is 
\begin{eqnarray}
X^{ab}_{\bb h} = X^{ab}_{\bb\alpha} = X^{ab\mu}_{\bb n} =
X^{ab\mu}_{\bb h_*} = 0 \,.
\end{eqnarray} 
This, together with the previous result rules out all terms
containing $\bb^a\lambda^B$.

Next are the terms containing $\bw^a_\mu\lambda^B$, which come from 
the variation of
\begin{eqnarray}
\bw^a_\mu\bw^b_\nu X^{ab\mu\nu}_{\bw_*\bw_*} + \bw^a_\mu\ba^b
X^{ab\mu}_{\bw_*\ba} + \bw^a_\mu h^b_\nu  X^{ab\mu\nu}_{\bw_* h_*}
+ \bw^a_\mu\partial_\nu n^b X^{ab\mu\nu}_{\bw_* n}
\end{eqnarray} 
As before, looking at the $\sr'$-variation of this I find that the
coefficients of $\bw^a_\mu h^b_\nu$, $\bw^a_\mu n^b$ and $\bw^a_\mu
\alpha^b$ imply 
\begin{eqnarray}
X^{ab\mu\nu}_{\bw_*\bw_*} = X^{ab\mu}_{\bw_*\ba} = 0\,, \qquad
X^{ab\mu\nu}_{\bw_* h_*} = \eta  X^{ab\mu\nu}_{\bw_* n} \,.
\end{eqnarray} 
Terms linear in $\bw^a_\mu$ come from varying 
\begin{eqnarray}
\bb^a X^a_{\bb} + \bw^a_\mu h^b  X^{ab\mu}_{\bw_* h} +
\bw^a_\mu\alpha^b  X^{ab\mu}_{\bw_* \alpha} + \bw^a_\mu \left( \eta
h^b_\nu + \partial_\nu n^b \right) X^{ab\mu\nu}_{\bw_* n}\,.
\end{eqnarray} 
The coefficient of $\bw^a_\nu$, or more precisely the functional
derivative $\delta_L/\delta\bw^a_\mu$, in the $\sr'$-variation of
this gives the relation 
\begin{eqnarray}
{1\over\zeta}\partial^\mu X^a_{\bb} - {1\over\xi}\Delta^b_R
X^{ab\mu}_{\bw_* h} -{1\over\zeta} \Delta'^b_R  X^{ab\mu}_{\bw_*
\alpha} + \left( {1\over\zeta}\partial_\nu f'^b - \Delta^b_{R\nu}
\right)  X^{ab\mu\nu}_{\bw_* n} = 0\,.
\label{a3.linomega}
\end{eqnarray} 
This equation contains $X^a_{\bb}$ which has non-vanishing mass
dimension, so it an contain fields other than $\theta^a$ and
$C^a_\mu$, and the argument used in previous cases cannot be
applied here. Therefore, the $X$'s appearing here must remain
undetermined for the moment.

Terms containing products of the form $\ba^a\lambda^B$ appear in the
$\sr'$-variation of 
\begin{eqnarray}
\ba^a\ba^b X^{ab}_{\ba\ba} + \ba^a h^b_\mu X^{ab\mu}_{\ba h_*} 
+ \ba^a\partial_\mu n^b X^{ab\mu}_{\ba n}
\end{eqnarray} 
The coefficients of $\ba^a h^b_\mu$ and $\ba^a\alpha^b$ in the
variation give 
\begin{eqnarray}
X^{ab}_{\ba\ba} = 0\,, \qquad  X^{ab\mu}_{\ba h_*} = \eta
X^{ab\mu}_{\ba n} \,.
\end{eqnarray} 
Terms linear in $\ba^a$ come from 
\begin{eqnarray}
\bb^a X^a_{\bb} - \ba^a h^b X^{ab}_{\ba h} + \ba^a\alpha^b
X^{ab}_{\ba\alpha} +  \ba^a\left( \eta h^b_\mu + 
\partial_\mu n^b \right) X^{ab\mu}_{\ba n}\,.
\end{eqnarray} 
The coefficient of $\ba^a$ in the $\sr'$-variation of this
satisfies
\begin{eqnarray}
X^a_{\bb} - {1\over\xi}\Delta^b_R  X^{ab}_{\ba h} -
{1\over\zeta}\Delta'^b_R X^{ab}_{\ba\alpha} + \left( \partial_\mu
f'^b - \Delta^b_{R\mu} \right) X^{ab\mu}_{\ba n} = 0\,.
\label{a3.linabar}
\end{eqnarray} 
This equation is again insufficient to determine the $X$'s in it
and will have to be reexamined later.

Terms containing $h^a\lambda^B$ come from the variation of 
\begin{eqnarray}
\bw^a_\mu h^b  X^{ab\mu}_{\bw_* h} + \ba^a h^b X^{ab}_{\ba h} +
h^ah^b_\mu X^{ab\mu}_{hh_*} + h^a\partial_\mu n^b X^{ab\mu}_{hn}\,, 
\end{eqnarray} 
the other possibilities being known to vanish from the above
analysis. The coefficients of $h^ah^b_\mu$, $h^an^b$ and
$h^a\alpha^b$ in the variation of this lead to 
\begin{eqnarray}
X^{ab\mu}_{\bw_* h} = X^{ab}_{\ba h} = 0\,, \qquad X^{ab\mu}_{hh_*}
= \eta  X^{ab\mu}_{hn}\,.
\end{eqnarray} 
Terms linear in $h^a$ come from 
\begin{eqnarray}
\bw^a X^a_{\bw} + h^a h^b X^{ab}_{hh} + h^a\alpha^b
X^{ab}_{h\alpha} + h^a\left( \eta h^b_\mu + \partial_\mu n^b
\right) X^{ab\mu}_{hn}\,.
\end{eqnarray} 
Of these, $X^{ab}_{h\alpha}$ has vanishing mass dimension and ghost
number $-1$. Since it is not possible to construct such a function
with the fields in the theory, it follows that $X^{ab}_{h\alpha} =
0$. The coefficient of $h^a$, in the terms linear in $h^a$, in the
variation of the rest satisfies the equation 
\begin{eqnarray}
- X^a_{\bw} + {2\over \xi}\Delta^b_R  X^{ab}_{hh} - \left
( {1\over\zeta} \partial_\mu f'^b - \Delta^b_{R\mu} \right)
X^{ab\mu}_{hn} = 0\,. 
\label{a3.linh}
\end{eqnarray} 
These terms will also be left for later scrutiny, as this equation
is insufficient to determine them.

Terms containing $\alpha^a\lambda^B$ come from the $\sr'$-variation
of 
\begin{eqnarray}
\bw^a_\mu \alpha^b X^{ab\mu}_{\bw_*\alpha} + \ba^a\alpha^b
X^{ab}_{\ba\alpha} + h^a_\mu h^b_\nu X^{ab\mu\nu}_{h_* h_*} +
h^a_\mu\partial_\nu n^b X^{ab\mu\nu}_{h_* n} + \partial_\mu
n^a\partial_\nu n^b X^{ab\mu\nu}_{nn}\,,
\end{eqnarray} 
where I have excluded terms that have already been shown to vanish,
and set 
\begin{eqnarray}
X^{ab}_{h\alpha} = X^{ab\mu}_{\alpha h_*} = X^{ab\mu}_{\alpha n} =
X^{ab}_{\alpha\alpha} = 0\,,
\end{eqnarray} 
because these $X$'s have negative ghost number and vanishing mass
dimension, so cannot be constructed out of fields present in the
theory. The terms containing products of the form
$\alpha^a\lambda^B$ in the $\sr'$-variation of of this can be set to
zero to give the equation
\begin{eqnarray}
-\left( h^a_\mu + {1\over\eta}\partial_\mu n^a \right) \alpha^b
X^{ab\mu}_{\bw_*\alpha} &-& {1\over\zeta}\partial_\mu \left( h^a_\mu
+ {2\over\eta}\partial_\mu n^a \right)\alpha^b X^{ab}_{\ba\alpha} 
- {2\over\eta}\partial_\mu\alpha^a h^b_\nu X^{ab\mu\nu}_{h_* h_*} 
\, \nonumber \\ 
&-& {1\over\eta} \partial_\mu\alpha^a \partial_\nu n^b
X^{ab\mu\nu}_{h_* n} + h^a_\mu \partial_\nu\alpha^b
X^{ab\mu\nu}_{h_* n} + 2  \partial_\mu\alpha^a \partial_\nu n^b
X^{ab\mu\nu}_{nn} = 0\,,
\label{a3.alphagamma}
\end{eqnarray} 
where I have used the fact that both $X^{ab\mu\nu}_{h_* h_*}$ and
$X^{ab\mu\nu}_{nn}$ are symmetric under the exchange $[\mu, a]
\leftrightarrow [\nu, b]$. This equation will also be put aside
for later use. Terms linear in $\alpha^a$ come from 
\begin{eqnarray}
\bw^a_\mu\alpha^b X^{ab\mu}_{\bw_*\alpha} + \ba^a\alpha^b
X^{ab}_{\ba\alpha} + h^a_\mu X^{a\mu}_{h_*} + \partial_\mu n^a
X^{a\mu}_{n} + \alpha^a X^a_\alpha \,.
\end{eqnarray} 
Of these, $X^a_\alpha$ has ghost number $-1$, and must satisfy
$\sr' X^a_\alpha = 0$ by definition (Eqn.(\ref{ren.srallX})), which
cannot happen unless $ X^a_\alpha = 0$. The terms linear in
$\alpha^a$ in the $\sr'$-variation of the rest lead to the equation 
\begin{eqnarray}
-{1\over\eta}f^b_\mu X^{ba\mu}_{\bw_*\alpha} - {2\over\zeta\eta}
\partial_\mu f^{b\mu} X^{ab}_{\ba\alpha} + {1\over\eta}\partial_\mu
X^{a\mu}_{h_*} - \partial_\mu X^{a\mu}_n = 0\,.
\label{a3.linalpha}
\end{eqnarray} 
The terms containing $h^a_\mu\lambda^B$ come from the
$\sr'$-variation of 
\begin{eqnarray}
\bw^a_\mu h^b X^{ab\mu}_{\bw_* h} &+& \bw^a_\mu\alpha^b
X^{ab\mu}_{\bw_*\alpha} + \left( \eta\,\bw^a_\mu h^b_\nu +
\bw^a_\mu\partial_\nu n^b \right) X^{ab\mu\nu}_{\bw_* n} +
\ba^a\alpha^b X^{ab}_{\ba\alpha} \nonumber \\ 
&+& \left( \eta\,\ba^a h^b_\mu + \ba^a\partial_\mu n^b \right)
X^{ab\mu}_{\ba n} + h^a_\mu h^b_\nu X^{ab\mu\nu}_{h_* h_*} +
h^a_\mu \partial_\nu n^b X^{ab\mu\nu}_{h_* n} \,.
\end{eqnarray} 
The coefficient of $h^a_\mu h^b$ in the variation of this gives $
X^{ab\mu}_{\bw_* h} = 0$, and the remaining terms with
$h^a_\mu\lambda^B$ in them satisfy
\begin{eqnarray}
-h^a_\mu\alpha^b X^{ab\mu}_{\bw_*\alpha} &-& \left( \eta\, h^a_\mu
h^b_\nu + 2 h^a_\mu \partial_\nu n^b \right) X^{ab\mu\nu}_{\bw_* n}
- {1\over\zeta}\, \partial_\mu h^{a\mu}\alpha^b X^{ab}_{\ba\alpha}
 - {2\over\eta}\,
\partial_\mu\alpha^a h^b_\nu X^{ab\mu\nu}_{h_* h_*}
\nonumber \\
&-&
\left[ {\eta\over\zeta}\,\partial_\nu \left(h^{a\nu} +
{2\over\eta}\partial^\nu n^a \right) h^b_\mu  +
{1\over\zeta}\,\partial_\nu h^{a\nu} \partial_\mu n^b \right]
X^{ab\mu}_{\ba n} +
h^a_\mu\partial_\nu\alpha^b X^{ab\mu\nu}_{h_* n} = 0 \,.
\end{eqnarray} 
It follows from this that $ X^{ab\mu\nu}_{\bw_* n} = X^{ab\mu}_{\ba
n} = 0$ (essentially because there is no linear combination of
$ba^a$ and $\partial_\mu\bw^a_\nu$ which is $\sr'$-invariant). Then
the remaining terms satisfy
\begin{eqnarray}
-h^a_\mu\alpha^b X^{ab\mu}_{\bw_*\alpha} - {1\over\zeta}\,
\partial_\mu h^{a\mu}\alpha^b X^{ab}_{\ba\alpha} - {2\over\eta}\,
\partial_\mu\alpha^a h^b_\nu X^{ab\mu\nu}_{h_* h_*} +
h^a_\mu\partial_\nu\alpha^b X^{ab\mu\nu}_{h_* n} = 0\,.
\label{ren.hgamma}
\end{eqnarray} 
Terms linear in $h^a_\mu$ appear from the $\sr'$-variation of 
\begin{eqnarray}
\bw^a_\mu X^{a\mu}_{\bw_*} + \ba^a X^a_{\ba} + h^a\left( \eta\,
h^b_\mu + \partial_\mu n^b \right) X^{ab\mu}_{hn} +  h^a_\mu
h^b_\nu X^{ab\mu\nu}_{h_* h_*} + h^a_\mu \partial_\nu n^b
X^{ab\mu\nu}_{h_* n}  \,,
\end{eqnarray} 
giving the following equation:
\begin{eqnarray}
-  X^{a\mu}_{\bw_*} + {1\over\zeta}\, \partial^\mu X^a_{\ba} + 
{\eta\over\xi}\,\Delta^b_R X^{ba\mu}_{hn} - {4\over\zeta\eta}
\partial_\nu f'^b  X^{ab\mu\nu}_{h_* h_*} + {2\over\eta}\,
\Delta^b_{R\nu}  X^{ab\mu\nu}_{h_* h_*} + {1\over\zeta}
\partial_\nu f'^b X^{ab\mu\nu}_{h_* n} = 0\,.
\label{a3.linhmu}
\end{eqnarray} 
Finally, terms containing $\partial_\mu n^a\lambda^B$ come from the
variation of 
\begin{eqnarray}
\bw^a_\mu\alpha^b X^{ab\mu}_{\bw_*\alpha} + \ba^a\alpha^b
X^{ab}_{\ba\alpha} + h^a_\mu \partial_\nu n^b X^{ab\mu\nu}_{h_* n}
+ \partial_\mu n^a \partial_\nu n^b X^{ab\mu\nu}_{nn} \,,  
\end{eqnarray} 
and satisfy the equation 
\begin{eqnarray}
- {1\over\eta}\,  \partial_\mu n^a\alpha^b X^{ab\mu}_{\bw_*\alpha}
- {2\over\zeta\eta}\Box n^a \alpha^b X^{ab}_{\ba\alpha} -
- {1\over\eta}\, \partial_\mu\alpha^a \partial_\nu n^b
X^{ab\mu\nu}_{h_* n} + 2 \partial_\mu n^a \partial_\nu\alpha^b
X^{ab\mu\nu}_{nn} = 0\,, 
\end{eqnarray} 
while terms linear in $n^a$ appear from the variation of 
\begin{eqnarray}
\bw^a_\mu X^{a\mu}_{\bw_*} + \ba^a X^a_{\ba} + h^a\partial_\mu n^b 
X^{ab\mu}_{hn} + h^a_\mu\partial_\nu n^b X^{ab\mu\nu}_{h_* n} + 
\partial_\mu n^a \partial_\nu n^b X^{ab\mu\nu}_{nn} \,,
\end{eqnarray} 
and gives the equation 
\begin{eqnarray}
-{1\over \eta}\,\partial_\mu n^a X^{a\mu}_{\bw_*} &-&
{2\over\zeta\eta}\, 
\Box n^a X^a_{\ba} + {1\over\xi}\Delta^a_R \partial_\mu n^b
X^{ab\mu}_{hn} \nonumber \\ &-&  {2\over\zeta\eta} \partial_\mu
f'^a \partial_\nu n^b  X^{ab\mu\nu}_{h_* n}
+ {1\over\eta}\Delta^a_{R\mu}\partial_\nu n^b  X^{ab\mu\nu}_{h_*
n} + {2\over\zeta} \partial_\mu f'^a \partial_\nu n^b
X^{ab\mu\nu}_{n n} = 0\,.
\end{eqnarray} 
There is one more equation that can be obtained from $\sr'\Gamma =
0$, the one involving terms which do not contain any of the
$\lambda^A$. This equation is
\begin{eqnarray}
-{1\over\xi}\,f^a X^a_{\bw} - {1\over \eta}\,f^a_\mu X^{a\mu}_{\bw_*} &-&
{2\over\zeta\eta}\,\partial_\mu f^a_\mu X^a_{\ba} +
{1\over\xi}\,\Delta^a_R X^a_h \nonumber \\ 
&-& {2\over\zeta\eta}\,\partial_\mu f'^a
X^{a\mu}_{h_*} + {1\over\eta}\,\Delta^a_{R\mu} X^{a\mu}_{h_*} +
{1\over\zeta}\,\partial_\mu f'^a  X^{a\mu}_n = 0\,.
\label{a3.nogamma}
\end{eqnarray} 

I can now write the effective Lagrangian for the ghost sector of
the theory, after setting to zero all the $X$'s that were found to
vanish in the analysis so far,
\begin{eqnarray}
{\mathscr L}_g = \bw^a X^a_{\bw} &+& \bb^a X^a_{\bb} + \bw^a_\mu
X^{a\mu}_{\bw_*} + \ba^a X^a_{\ba} + h^a X^a_h + h^a_\mu
X^{a\mu}_{h_*} + \partial_\mu n^a X^{a\mu}_n \, \nonumber \\
&+&  \bw^a_\mu \alpha^b X^{ab\mu}_{\bw_*\alpha} + \ba^a \alpha^b
X^{ab}_{\ba\alpha} + h^a h^b X^{ab}_{hh} + h^a \left( \eta h^b_\mu
+ \partial_\mu n^b \right) X^{ab\mu}_{hn} \,\nonumber \\
&+& h^a_\mu h^b_\nu
X^{ab\mu\nu}_{h_* h_*} + h^a_\mu\partial_\nu n^b X^{ab\mu\nu}_{h_*
n} + \partial_\mu n^a \partial_\nu n^b X^{ab\mu\nu}_{nn}\,.
\label{a3.ghostac}
\end{eqnarray} 
%

\section{Derivation of Eqn.~(\ref{LGFINAL})}\label{a4}
In this Appendix, I shall try to calculate the functions which
remained undetermined in Eqn.~(\ref{SR'GAMMA}) (equivalently,
Eqn.~(\ref{a3.ghostac})). I start from the expression for the
$\sr$-variation of $\Gamma$, which can be written as
\begin{eqnarray}
\sr\Gamma = && \sum\limits_A^{}\left[ (\sr\lambda^A)X^A +
(-1)^{\varepsilon_A} \lambda^A(\sr X^A) \right] \,\nonumber \\
&&+ \sum\limits_{AB}^{}
\left[ (\sr\lambda^A)\lambda^B X^{AB} + (-1)^{\varepsilon_A}\lambda^A
(\sr\lambda^B) X^{AB} + (-1)^{\varepsilon_A + \varepsilon_B}\lambda^A
\lambda^B (\sr X^{AB}) \right] = 0.
\end{eqnarray} 
Since $\sr X^A$ and $\sr X^{AB}$ do not contain any of the
$\lambda^A$, I can consider the coefficients of $\lambda^A$ or of
$\lambda^A\lambda^B$ in the above expression and set them to zero
one at a time. I will first set to zero the coefficients of
$\lambda^A\lambda^B$, and then the terms linear in $\lambda^A$.

The coefficient of $\partial_\mu n^a \partial_\nu\alpha^b$ and 
$h^a\partial_\mu\alpha^b$ give
\begin{eqnarray}
X^{ab\mu\nu}_{nn} = X^{ab\mu}_{hn} = 0 \,,
\end{eqnarray} 
while the coefficients of $\ba^a\alpha^b, h^a h^b, h^a_\mu h^b_\nu$ and
$h^a_\mu\partial_\nu n^b$ give 
\begin{eqnarray}
\sr X^{ab}_{\ba\alpha} =  \sr X^{ab}_{hh} = \sr X^{ab\mu\nu}_{h_* h_*} 
= \sr X^{ab\mu\nu}_{h_* n} = 0\,.
\label{ren.srX}
\end{eqnarray} 
Now, each of the $X$'s in this equation has zero mass dimension,
zero ghost number and is $\sr$-invariant, so each must be a
(possibly different) constant.  Let me define four constants $K_1,
K_2, K_3$ and $K_4$ as
\begin{eqnarray}
X^{ab}_{\ba\alpha} = K_1 \delta^{ab} \,,\qquad  X^{ab}_{hh} = K_2
\delta^{ab} \,, \qquad  X^{ab\mu\nu}_{h_* h_*} = K_3 g^{\mu\nu}
\delta^{ab} \,, \qquad X^{ab\mu\nu}_{h_* n} = K_4^{\mu\nu}
\delta^{ab} \,. 
\end{eqnarray} 
The coefficients of $h^a, \ba^a, h^a_\mu, \alpha^a,
\bw^a, \bb^a, \bw^a_\mu$ and $n^a$ give the equations
\begin{eqnarray}
- X^a_{\bw} + \sr X^a_h &=& 0 \,, \nonumber \\
X^a_{\bb} - \sr X^a_{\ba} &=& 0 \,, \nonumber \\
- X^{a\mu}_{\bw_*} + \sr X^{a\mu}_{h_*} &=& 0 \,, \nonumber \\
\partial_\mu X^{a\mu}_n &=& 0 \,, \nonumber \\
\sr X^a_{\bw} = \sr X^a_{\bb} = \sr X^a_{\bw_*} &=& \sr \partial_\mu
X^{a\mu}_n = 0 \,. 
\label{ren.Xeqns}
\end{eqnarray} 
The last equation in this list is redundant as it can be obtained by
applying $\sr$ to the previous equations and remembering that $\sr^2 =
0$.The coefficient of $\bw^a_\mu\alpha^b$ gives $\sr
X^{ab\mu}_{\bw_*\alpha} = 0$, but $X^{ab\mu}_{\bw_*\alpha}$ must
contain a derivative operator to allow for the constant shift
symmetry of $\bw^a_\mu$. So as with the functions in
Eqn.(\ref{ren.srX}), $X^{ab\mu}_{\bw_*\alpha}$ must be a constant
times a derivative operator. The coefficient of $h^a_\mu\alpha^b$
in $\sr {\mathscr L}_g$ shows that this constant is $ K_4^{\mu\nu}$,
so I can write the ghost sector Lagrangian as
\begin{eqnarray}
{\mathscr L}_g = \bw^a X^a_{\bw} &+& \bb^a X^a_{\bb} + \bw^a_\mu
X^{a\mu}_{\bw_*} + \ba^a X^a_{\ba} + h^a X^a_h + h^a_\mu
X^{a\mu}_{h_*}  \,
\nonumber \\ 
&-& K^{\mu\nu}_4 \partial_\nu\bw^a_\mu \alpha^a + K_1\ba^a
\alpha^a + K_2 h^a h^a + K_3 h^a_\mu h^a_\mu + 
K^{\mu\nu}_4 h^a_\mu\partial_\nu n^a \,.
\label{ren.Lgmid}
\end{eqnarray} 
Now I can use the unused equations from Appendix~\ref{a3}. These
were Eqs.~(\ref{a3.linomega}), (\ref{a3.linabar}), (\ref{a3.linh}),
(\ref{a3.alphagamma}), (\ref{a3.linalpha}), (\ref{a3.linhmu}), and
(\ref{a3.nogamma}).

Eqn.(\ref{a3.linabar}) now reads 
\begin{eqnarray}
X^a_{\bb} - {K_1\over\zeta}\Delta'^a_R = 0 \,.
\end{eqnarray} 
In keeping with standard notation, let me rewrite 
\begin{eqnarray}
K_1 = \zeta Z_\beta \, \qquad \Rightarrow \qquad X^a_{\bb} =
Z_\beta \Delta'^a_R \,.
\label{ren.Xbeta}
\end{eqnarray} 
Eqn.(\ref{a3.linh}) now becomes
\begin{eqnarray}
- X^a_{\bw} + {2\over \xi} K_2 \Delta^a_R = 0 \,.
\end{eqnarray} 
As in the above, let me redefine the constant,
\begin{eqnarray}
K_2 = {\xi\over 2}Z_\omega \qquad \Rightarrow \qquad 
X^a_{\bw} = Z_\omega \Delta^a_R \,.
\label{ren.Xomega}
\end{eqnarray} 
Eqn.(\ref{a3.linomega}) becomes
\begin{eqnarray}
{1\over \zeta}\partial^\mu X^a_{\bb} - {1\over \zeta} K_4^{\mu\nu}
\partial_\nu \Delta'^a_R = 0 \,,
\end{eqnarray} 
which gives upon using Eqn.(\ref{ren.Xbeta}) that 
\begin{eqnarray}
K_4^{\mu\nu} = Z_\beta\, g^{\mu\nu} \,.
\end{eqnarray} 
Using this and Eqn.(\ref{ren.Xbeta}) I can rewrite
Eqn.(\ref{ren.hgamma}) as 
\begin{eqnarray}
Z_\beta\, \partial_\mu h^{a\mu}\alpha^a - Z_\beta\, 
\partial_\mu h^{a\mu}\alpha^a - {2\over \eta}\, K_3\, h^{a\mu}
\partial_\mu \alpha^a + Z_\beta\, h^{a\mu}\partial_\mu
\alpha^a = 0 \,,
\end{eqnarray} 
from which it follows that 
\begin{eqnarray}
K_3 = {\eta\over 2}Z_\beta \,.
\end{eqnarray} 
This automatically satisfies Eqn.(\ref{a3.alphagamma}). With these
redefinitions I get from Eqn.(\ref{a3.linalpha}) that
\begin{eqnarray}
\partial_\mu X^{a\mu}_{h_*} = Z_\beta \partial_\mu
f^{a\mu}\,.
\label{ren.Xhmu}
\end{eqnarray} 
The right hand side vanishes upon using $f^{a\mu} = \partial_\nu
B^{a\mu\nu}$. Also, Eqn.(\ref{a3.linhmu}) can be written as
\begin{eqnarray}
- X^{a\mu}_{\bw_*} + {1\over\zeta}\,\partial^\mu X^a_{\ba} -
{1\over\zeta}\, Z_\beta\, \partial^\mu f'^a  
+ Z_\beta\,\Delta^{a\mu}_R = 0 .
\label{ren.Xomegamu}
\end{eqnarray} 
Using
Eqs.~(\ref{ren.Xeqns}),(\ref{ren.Xbeta}),(\ref{ren.Xomega}) and 
(\ref{ren.Xomegamu}), 
I can define some new functions and write 
\begin{eqnarray}
X^a_h &=& Z_\omega f^a + \bar X^a_h \,,\qquad  X^a_{\ba} =
Z_\beta f'^a + \bar X^a_{\ba} \,, \nonumber \\
X^{a\mu}_{\bw_*} &=& 
Z_\beta \Delta^{a\mu} + {1\over \zeta}\, \partial^\mu \bar
X^a_{\ba} \,,\qquad X^{a\mu}_{h_*} = Z_\beta f^{a\mu} +
 {1\over \zeta}\,\partial^\mu \bar X^a_{h_*} \,, 
\qquad \sr \bar X^a_{h_*} =  \bar X^a_{\ba} \,.
\end{eqnarray} 
Then Eqn.(\ref{ren.Xhmu}) implies, because $ \bar X^a_{h_*}$ is a
function of the fields and not an arbitrarily chosen function, that
\begin{eqnarray}
\bar X^a_{h_*} = 0\,,\; {\rm and \; hence\;} \bar X^a_{\ba} = 0 \,.
\end{eqnarray} 
Putting these into Eqn.(\ref{a3.nogamma}), I get $\bar X^a_h =
0$. Therefore, I can now write down the general form of the ghost
sector of the theory as
\begin{eqnarray}
{\mathscr L}_g = Z_\omega\, \bw^a \Delta^a_R &+& 
Z_\beta\, \bb^a \Delta'^a_R + Z_\beta\, \bw^a_\mu
\Delta^{a\mu}_R + Z_\beta\, \ba^a f'^a + Z_\omega\,
h^a f^a + Z_\beta\, h^a_\mu \left( f^{a\mu} + \partial^\mu
n^a \right) \nonumber \\
&-& Z_\beta\, \partial^\mu\bw^a_\mu \alpha^a + \zeta\, 
Z_\beta\, \ba^a\alpha^a + {\xi\over 2}\,Z_\omega\, h^ah^a  
+ {\eta\over 2}\,Z_\beta\, h^a_\mu h^{a\mu}\,.
\label{a3.lgfinal}
\end{eqnarray} 
%



\begin{table}
\caption{Mass dimensions and ghost numbers of the fields and their
antisources.\label{a3.t.mdgn}} 
\begin{center}
\begin{tabular}{lcccc}
\hline
Field & dimension & ghost number & dimension &
ghost number \\  
$\chi^A$ & & &  of $K^A$ &  of $K^A$ \\ \hline
$A^a_\mu$ & 1 & 0 & 2 & $-1$ \\ 
$B^a_{\mu\nu}$ & 1 & 0 & 2 & $-1$ \\ 
$C^a_\mu$ & 0 & 0 & 3 & $-1$ \\ 
$\omega^a$ & 1 & 1 & 2 & $-2$ \\ 
$\bw^a$ & 1 & $-1$ & 2 & 0 \\ 
$h^a$ & 2 & 0 & $\star$ & $\star$ \\ 
$\omega^{a\mu}$ & 1 & 1 & 2 & $-2$ \\ 
$\bw^{a\mu}$ & 1 & $-1$ & 2 & 0 \\ 
$\theta^{a}$ & 0 & 1 & 3 & $-2$ \\ 
$h^{a\mu}$ & 2 & 0 & $\star$ & $\star$ \\ 
$n^{a}$ & 1 & 0 & 2 & $-1$ \\ 
$\beta^{a}$ & 1 & 2 & 2 & $-3$ \\ 
$\bb^{a}$ & 1 & $-2$ & 2 & 1 \\ 
$\alpha^{a}$ & 2 & 1 & $\star$ & $\star$ \\ 
$\ba^{a}$ & 2 & $-1$ & $\star$ & $\star$ \\ 
$\alpha$ & 0 & 1 & $\star$ & $\star$ \\ 
\hline
\end{tabular}
\end{center}
\end{table}
A $\star$ indicates that the antisource $K^A$ does not appear in the
theory as the BRST variation of the corresponding field $\chi^A$
vanishes.


\begin{thebibliography}{99}
\baselineskip=23pt

\bibitem{thooft}{G. 't Hooft, {\sl Nucl. Phys.} {\bf B35}, (1971) 167.}
\bibitem{aurtak}{A. Aurilia and Y. Takahashi, {\sl Prog. Theor. Phys.}
{\bf 66}, (1981) 693.}
\bibitem{trg}{T. R. Govindarajan, {\sl J. Phys. G} {\bf 8}, (1982) L17.}
\bibitem{abl}{T. J. Allen, M. J. Bowick and A. Lahiri,
{\sl Mod. Phys. Lett.} {\bf A6}, (1991) 559.}
\bibitem{minwar}{J. A. Minahan and R. C. Warner, {\sl St\" uckelberg
Revisited}, Florida U. Preprint UFIFT-HEP-89-15.}
\bibitem{hurth}{T.~Hurth, {\sl Helv. Phys. Acta} {\bf 70}, (1997) 406,
{\tt hep-th/9511176.}}
\bibitem{nogo}{M.~Henneaux et al, {\sl Phys. Lett.} {\bf B410}, (1997)
195, {\tt hep-th/9707129.}}
\bibitem{gvm} {A.~Lahiri, {\sl Generating vector boson masses},
Los Alamos report no. LA-UR-92-3477, {\tt hep-th/9301060}.}
\bibitem{hwalee} {D.~S.~Hwang and C.~-Y.~Lee, {\sl J. Math. Phys.} 
{\bf 38}, 30 (1997), {\tt hep-th/9512216}.}
\bibitem{nabrst} A.~Lahiri, {\sl Phys. Rev.} {\bf D55},  (1997) 5045, 
{\tt hep-ph/9609510}.
\bibitem{neto} {J.~Barcelos-Neto and S.~Rabello, {\sl Z.Phys.} {\bf
C74}, (1997) 715-719, {\tt hep-th/9601076.}}
\bibitem{ZJ} {J.~Zinn-Justin, {\sl Quantum Field Theory and Critical
Phenomena}, Clarendon Press, Oxford, 1989.}
\bibitem{weinqft} {S.~Weinberg, {\sl The Quantum Theory of Fields, 
Vol. 2: Modern Applications}, Cambridge University Press, 1996.}
\bibitem{piso} {O.~Piguet and S.~P.~Sorella, {\sl Algebraic
Renormalization}, Springer Lecture Notes in Physics vol. 28, 1995.}
\bibitem{hentei} {M.~Henneaux and C.~Teitelboim, {\sl Quantization of
Gauge Systems}, Princeton University Press, 1992.}
\bibitem{catzen} {A.~S.~Cattaneo et al, {\sl Commun. Math. Phys.} {\bf
197}, (1998) 571, {\tt hep-th/9705123}.}
\bibitem{nickm}{N.~S.~Manton, {\sl J.Math.Phys.} {\bf 40}, (1999)
736, {\tt hep-th/9806077}.}


\end{thebibliography}
\end{document}